\documentclass[11pt]{article}

\usepackage{array}
\usepackage{graphicx}
\usepackage{amsmath}
\usepackage{hhline}

\title{\bf Conjugate Gradient Method for finding fundamental solitary waves}

\author{ T.I. Lakoba\footnote{lakobati@cems.uvm.edu, \ 1 (802) 656-2610} 
 \vspace{1cm} \\
  Department of Mathematics and Statistics, 16 Colchester Ave., \\
 University of Vermont, Burlington, VT 05401, USA}

\newcommand{\noi}{\noindent}
\newcommand{\D}{\Delta}
\newcommand{\be}{\begin{equation}}
\newcommand{\ee}{\end{equation}}
\newcommand{\bsube}{\begin{subequations}}
\newcommand{\esube}{\end{subequations}}
\newcommand{\ba}{\begin{array}}
\newcommand{\ea}{\end{array}}
\newcommand{\To}{\rightarrow}
\newcommand{\vecx}{{\bf x}}
\newcommand{\vecy}{{\bf y}}
\newcommand{\vecz}{{\bf z}}
\newcommand{\vecb}{{\bf b}}
\newcommand{\vecr}{{\bf r}}
\newcommand{\vecd}{{\bf d}}
\newcommand{\PM}{Petviashvili method}

\newcommand{\Lnot}{L^{(0)}}
\newcommand{\Lnnot}{L^{(00)}}
\newcommand{\Lcal}{{\mathcal L}}
\newcommand{\Lcalnot}{{\mathcal L}^{(0)}}
\newcommand{\Knot}{K^{(0)}}
\newcommand{\Knnot}{K^{(00)}}
\newcommand{\Kcal}{{\mathcal K}}
\newcommand{\Kcalnot}{{\mathcal K}^{(0)}}
\newcommand{\vecQ}{\vec{Q}}
\newcommand{\vecmu}{\vec{\mu}}
\newcommand{\Ucal}{{\mathcal U}}

\newcommand{\bea}{\begin{eqnarray}}
\newcommand{\eea}{\end{eqnarray}}
\newcommand{\so}{\Rightarrow}
\newcommand{\dst}{\displaystyle}

\newcommand{\tu}{\tilde{u}}
\newcommand{\tv}{\tilde{v}}

\newcommand{\vecN}{{\bf N}}

\newcommand{\bptrans}{\bar{\psi}_{\rm trans}}
\newcommand{\bpperp}{\bar{\psi}_{\perp}}

\renewcommand{\theequation}{\thesection.\arabic{equation}}

\begin{document}
\baselineskip 18 pt

\maketitle

\vspace*{3cm}

\begin{center}
 {\bf Abstract}
\end{center}

The Conjugate Gradient method (CGM) is known to be the fastest generic iterative
method for solving linear systems with symmetric sign definite matrices. In this
paper, we modify this method so that it could find fundamental solitary waves of
nonlinear Hamiltonian equations. The main obstacle that such a modified CGM overcomes
is that the operator of the equation linearized about a solitary wave is not sign
definite. Instead, it has a finite number of eigenvalues on the opposite side of zero
than the rest of its spectrum. We present versions of the modified CGM 
that can find solitary waves with prescribed values of either the propagation constant 
or power. We also extend these methods to handle multi-component nonlinear wave equations. 
Convergence conditions of the proposed methods are given, and their practical
implications are discussed. We demonstrate that our modified CGMs converge much
faster than, say, Petviashvili's or similar methods, especially when the latter
converge slowly.

\vskip 1.1 cm

\noi

\bigskip

\noi

\newpage

\section{Introduction}

Solitary waves of most nonlinear evolution equations can be found only numerically. 
In one spatial dimension, a variety of numerical methods for finding solitary waves are
available, with the shooting and Newton's methods being probably the most widely used
ones. However, in two and three spatial dimensions, 
the shooting
method is no longer applicable, and the complexity of programming of Newton's method 
increases considerably compared to the one-dimensional case. 
Also, the accuracy of
Newton's method is only algebraic in the size of the spatial discretization step 
$ \D x$ (typically, it is 
$O\left( (\D x)^2\right)$). This accuracy may be too low for some applications, since in 
two or three spatial dimensions, one tends to take larger discretization steps than in
one dimension in order to limit the computational time and the size of stored numeric
arrays. Therefore, it is desirable to have a method that would: \ (i) be easier to program
than Newton's method; \ (ii) have the exponential accuracy in $\D x$, as spectral methods;
\ and \ (iii) converge sufficiently fast under known conditions.

A number of such methods are available. The well-known iterative method
proposed by Petviashvili \cite{Petviashvili76} can be used
to find stationary solitary waves of equations with power-law nonlinearity:
\be
-Mu+u^p=0\,, \qquad u(\vecx)\To 0 \quad {\rm as} \quad |\vecx|\To\infty,
\label{e1_01}
\ee
where $u$ is the real-valued field of the solitary wave, $M$ is a
positive definite and self-adjoint differential operator with
constant coefficients, and $p$ is a constant. For example, the
solitary wave of the nonlinear Schr\"odinger equation in $D$ spatial
dimensions,
\be
\ba{cc}
iU_t+\nabla^2 U + |U|^2U=0\,, & \quad U(|\vecx|\To \infty) \To 0\,, \vspace{0.1cm} \\
\displaystyle  \nabla^2 \equiv \frac{\partial^2}{\partial x_1^2} + \cdots
            + \frac{\partial^2}{\partial x_D^2} \,, &
\ea
\label{e1_02}
\ee
upon the substitution \
$ U(\vecx,t)=e^{i\mu t}u(\vecx)$ \ 
reduces to Eq. (\ref{e1_01}) with $p=3$ and 
\be
M=\mu-\nabla^2\,.
\label{e1_03}
\ee
The parameter $\mu>0$ is referred to as the propagation constant of the solitary wave.
Convergence conditions of the \PM\ for Eq.~(\ref{e1_01}) were established in \
\cite{PelinovskyS04}.

Studies of solitary waves in nonlinear photonic lattices and Bose--Einstein condensates
motivated the interest in
finding solitary waves of equations that have a more general form than 
(\ref{e1_01}):
\be
-Mu+F(u,\vecx)=0\,, \qquad u(\vecx)\To 0 \quad {\rm as} \quad |\vecx|\To\infty,
\label{e1_04}
\ee
where $F(u,\vecx)$ is a real function.
For example, solitary waves in nonlinear photonic lattices with Kerr nonlinearity
satisfy the following equation of the form (\ref{e1_05}):
\be
\nabla^2 u + V_0(\cos^2x + \cos^2y) \,u + u^3\,=\,\mu u\,,
\label{e1_05}
\ee
where the second term accounts for the effect of the periodic potential provided
by the lattice.
Various modifications of the \PM\ for obtaining solutions of (\ref{e1_04})
were proposed; see, e.g., \cite{MusslimaniY04,YangM04,AblowitzM05,StepanyantsT06}. 
However, convergence conditions of those modifications were not studied.
In Ref.~\cite{gP,SOM}, J. Yang and the present author proposed two alternative 
iterative methods
which satisfied conditions (i)--(iii) stated before Eq.~(\ref{e1_01}); in particular,
the conditions of their convergence 
to fundamental (see below) \cite{gP} and 
both fundamental and non-fundamental
solitary waves (also known as ground and excited states, respectively) \cite{SOM}
were given. In Refs. \cite{ME,SOM}, a technique referred there as mode elimination
was proposed, which was shown to provide considerable acceleration of iterative methods
when the later converged slowly to the respective solitary wave.

In this paper, we propose yet another family of numerical methods for finding 
fundamental solitary waves of Hamiltonian nonlinear evolution equations.
Consider a linear system
\be
A\vecy - \vecb= {\bf 0},
\label{e1_06}
\ee
for the unknown vector $\vecy$, where  $A$ is a real symmetric matrix and $\vecb$ is a known vector. 
Then the methods proposed here are 
extensions of the well-known Conjugate Gradient method (CGM) for
the linear system \eqref{e1_06}
to the nonlinear boundary-value problem (\ref{e1_04}), and they
converge even faster than, e.g., the
generalized \PM\  \cite{gP} accelerated by mode elimination \cite{ME}.

The main part of this paper is organized as follows. In Section 2, we
explain how the methods introduced in \cite{gP,SOM,ME} for the nonlinear
problem (\ref{e1_04}) are related to some
well-known iterative  methods of solving the linear system \eqref{e1_06} 
with a real and symmetric matrix $A$. This comparative review 
will lead us to explaining the main issue that needs to be resolved in
order to extend the CGM in such a way that it could find fundamental solitary waves.
In Section 3, we present such an extension of the CGM for the case where
the solitary wave of the single-component 
Eq.~(\ref{e1_04}) has a prescribed value of the propagation constant $\mu$.
In Section 4, we present a modification of the CGM for finding fundamental
waves of (\ref{e1_04}) with a prescribed value not of the propagation constant
but of power, defined as 
\be
P=\int u^2(\vecx)\,d\vecx\,,
\label{e1_27}
\ee
where the integration is over the entire spatial domain. 
Let us note that another method for finding solitary waves with a specified value of power 
was previously considered in, e.g., \cite{GarciaRipollPG01,BaoD03,ShchesnovichC04,ITEM},
with its convergence conditions for \eqref{e1_04} being established in \cite{ITEM}. 
We will refer to this method as the imaginary-time evolution method.
In Section 5, we generalize the modified CGMs of Sections 3 and 4 to 
multi-component equations. In Section 6, we compare the performances of the generalized
Petviashvili and imaginary-time evolution methods
accelerated by the mode elimination technique \cite{ME,SOM} 
with the performances of the respective modified CGMs from Sections 3--5.
In Section 7 we summarize this work.
Appendix 1 contains proofs and discussions of some auxiliary results of Sections 3 and 4.
Appendix 2 contains an alternative method to the modified CGM of Section 3.
Appendix 3 contains a sample code of a modified CGM for a two-component 
equation considered in Section 6.

Before concluding this Introduction, we need to discuss what we will refer to
as a {\em fundamental solitary wave}. In fact, we are not aware of a universal definition
of this term. In many situations, one can intuitively identify a solitary wave as being
fundamental if it satisfies two easily-verifiable conditions. 
First, it is to have one main peak, with smaller peaks possibly existing around 
the main one. Second, for envelope solitary waves (e.g., those satisfying the complex-valued
Eq. \eqref{e1_02}), the propagation constant of the 
fundamental wave must lie in the semi-infinite spectral bandgap; see Fig.~\ref{fig_1}
in Section 6. (Typically, only equations with periodic potentials or systems
describing more than two linearly coupled waves propagating with different group 
velocities have more than one
spectral bandgap.) Alternatively, for carrier (e.g., Korteweg--de Vries-type) solitary waves, 
the velocity, rather than the propagation constant, of a fundamental wave
must lie in the semi-infinite gap. More rigorous characterization of fundamental waves
is possible in special cases. For instance, when the spatial operator in the wave equation is
the Laplacian (as, e.g., in \eqref{e1_02} or \eqref{e1_05}), the fundamental wave
is known to be nonzero for all $\vecx$ (see, e.g., \cite{Struwe00}); however, for a more general
operator as, e.g., in the Kadomtsev--Petviashvili equation, the fundamental wave may
have zeros \cite{KPlump}. For some equations (see, e.g., \cite{Sulem_book,Liu07}), 
it was rigorously shown that the fundamental solution minimizes a functional known as
the action. 

The property of an $S$-component fundamental solitary wave that {\em we will rely on 
in this paper} is that for many such waves, the linearized operator of the stationary equation 
has no more than $S$ positive eigenvalues. 
To the author's knowledge, there is no general proof
of this property even for a single-component Eq. \eqref{e1_04}, let alone for the $S$-component
case. (See, however, a recent review of solitary wave stability in lattices \cite{Sivan08}.)
Therefore, we simply {\em declare this property to be our working definition of a fundamental solitary wave}
for the purpose of this paper. Again, we do {\em not} imply that any solitary wave that has the
two intuitive properties described at the beginning of the previous paragraph also satisfies the
above property about the positive eigenvalues of the corresponding linearized operator.
Rather, we only say that it is only for the waves that do have the latter property that the
numerical methods developed in this paper will be guaranteed to converge (with possible restrictions,
as described in subsequent sections). 
For solitary waves whose corresponding linearized operators have more positive eigenvalues than
the number of the wave's component, our modified CGMs are not guaranteed to converge.

\section{Comparative review of iterative methods for nonlinear problem (\ref{e1_04})
and for linear problem (\ref{e1_06})}
\setcounter{equation}{0}

The reader who is not interested in such a review and wants to see the description
of new algorithms proposed in this paper may skip to Section 3.

For reference purposes, let us rewrite (\ref{e1_04}) as
\be
L^{(0)}u=0\,.
\label{e1_07}
\ee
An important role in our analysis will be played by the linearized operator $L$ of
the nonlinear Eq.~(\ref{e1_07}). Let $u$ be the exact solution of (\ref{e1_07}),
$u_n$ be the approximation of that solution obtained at the $n$th iteration, and
\be
u_n=u+\tu_n, \qquad \mbox{where $|\tu_n|\ll |u|$.}
\label{e1_08}
\ee
Then
\be
\Lnot u_n= \Lnot u+L\tu_n + O(|\tu_n|^2)=L\tu_n + O(|\tu_n|^2)\,.
\label{e1_09}
\ee
For example, for the nonlinear equation (\ref{e1_05}), the linearized operator is
\be
L=-M + V_0(\cos^2x + \cos^2y)+3u^2\,.
\label{e1_10}
\ee
In our discussion, the linear system (\ref{e1_06}) is a counterpart of the nonlinear
equation (\ref{e1_07}) and matrix $A$ is the counterpart of the linearized operator $L$.
For Hamiltonian wave equations that give rise to (\ref{e1_07}), operator $L$ is always
self-adjoint, which is the counterpart of matrix $A$ being real and symmetric. 
Having stated this correspondence between the nonlinear problem (\ref{e1_07}) and its
linear counterpart (\ref{e1_06}), we will, from now on, refer to them interchangeably,
because our statements will apply to both of them equally.

As we will explain below, the linearized forms of the iterative methods proposed
in \cite{Petviashvili76,gP,SOM} for solving the nonlinear problems (\ref{e1_01})
and (\ref{e1_04}) are related to Richardson's method\footnote{
also known as Picard's or fixed-point iterative method}
(see, e.g., \cite{TrefethenBau_NLA}, p. 274)
of solving the linear problem (\ref{e1_06}):
\be
\vecy_{n+1}=\vecy_n+(A\vecy_n-\vecb)\D\tau\,,
\label{e1_11}
\ee
The necessary condition for the iteration scheme (\ref{e1_11}) to converge to
the solution, $\vecy=A^{-1}\vecb$,
is that $A$ be negative definite. Choosing the parameter $\D\tau$ in (\ref{e1_11})
so that \ $\D\tau < -2/\lambda_{\min}$, \ where $\lambda_{\min}$ is the minimum
(i.e., most negative) eigenvalue of $A$, one ensures that Richardson's method
for (\ref{e1_06}) with a negative definite $A$ converges.

A naive counterpart of (\ref{e1_11}) for the nonlinear problem (\ref{e1_07}) is
\be
u_{n+1}=u_n+ \Lnot u_n\,\D\tau\,.
\label{e1_12}
\ee
However, in most cases it will not converge to a solitary wave. Indeed, 
from (\ref{e1_09}), the linearized form of (\ref{e1_12}) is
\be
\tu_{n+1}=\tu_n+L\tu_n\D\tau\,.
\label{e1_13}
\ee
This equation is a counterpart of (\ref{e1_11}), and so a necessary condition 
for its convergence is that operator $L$ be negative definite. However, 
for most fundamental single-component solitary waves,
the corresponding linearized operator $L$ has one positive eigenvalue
(see the last paragraph of the Introduction), which causes divergence of small deviations $\tu_n$
in (\ref{e1_13}) and, thereby, divergence of the iterations (\ref{e1_12}).

To overcome this divergence, the iterative methods of \cite{Petviashvili76,gP,SOM}
replaced $\Lnot u_n$ in (\ref{e1_12}) by a modified expression $L^{(0,\rm mod)}u_n$
such that: \ (i) \ 
$\left( \Lnot u=0 \right) \Leftrightarrow \left( L^{(0,\rm mod)} u=0 \right) $ \ 
and \ (ii) \ the corresponding linearized operator $L^{(\rm mod)}$ was guaranteed
to be negative definite. For example, in \cite{SOM}, 
\be
L^{(0,\rm mod)} u = -L\,\Lnot u,
\label{e1_14}
\ee
so that $L^{(\rm mod)}=-L^2$. In terms of the
linear system (\ref{e1_06}) where $A$ is real symmetric but not sign definite,
this is equivalent to applying Richardson's method (\ref{e1_11}) to the so called
normal equation 
\be
-A^2\vecy = -A\vecb,
\label{e1_15}
\ee
where the matrix on the left-hand side is always negative definite.

In \cite{gP}, we proposed an alternative idea of constructing a negative definite 
$L^{(\rm mod)}$. This idea will play an important role in the development of this paper,
and here we will explain it as applied to the linear system (\ref{e1_06}).
As noted above, for most single-component fundamental solitary waves, the linearized operator $L$ has
only one positive eigenvalue. Accordingly, let $\lambda^{(1)}$ be the only positive 
eigenvalue of matrix $A$ in (\ref{e1_06}), with the corresponding eigenvector being 
$\vecz^{(1)}$. In the context of the nonlinear equation (\ref{e1_07}), the eigenvector
of $L$ corresponding to its only positive eigenvalue
is related to the solution $u$ of (\ref{e1_07}) \cite{gP}, as we will explain later.
Hence it can be
considered as approximately known once the iterative solution $u_n$ is sufficiently
close to $u$, which we will always assume to be the case (see (\ref{e1_08})). 
Now, instead of (\ref{e1_06}), consider an {\em equivalent} system
\be
\left( I-\gamma \vecz^{(1)} \left( \vecz^{(1)}\right)^T \right)
\left( A\vecy - \vecb \right) = {\bf 0}\,,
\label{e1_16}
\ee
where $I$ is the identity matrix and $\gamma$ is any number greater than one. 
Note that the term \ $\vecz^{(1)} \left( \vecz^{(1)}\right)^T$ \ is the projection
matrix onto the unstable direction $\vecz^{(1)}$.
Using the
spectral decomposition of a real symmetric $p\times p$ matrix $A$:
\be
A= \sum_{i=1}^p \lambda^{(i)} \vecz^{(i)} \left( \vecz^{(i)}\right)^T \,,
\label{e1_17}
\ee
where the eigenvectors $\vecz^{(i)}$ form an orthonormal set:
\be
\left(\vecz^{(i)}\right)^T \vecz^{(j)} = 
\left\{ \ba{ll} 1, & i=j \\ 0, & i \neq j, \ea \right.
\label{e1_18}
\ee
it is straightforward to show that (\ref{e1_16}) and hence (\ref{e1_06}) is equivalent to 
\bsube
\be
A^{(\rm mod)} \vecy - \left( I-\gamma \vecz^{(1)} \left( \vecz^{(1)}\right)^T \right)
\vecb = {\bf 0}, 
\label{e1_19a}
\ee
where
\be
A^{(\rm mod)} \equiv \lambda^{(1)}(1-\gamma) \vecz^{(1)} \left( \vecz^{(1)}\right)^T +
\sum_{i=2}^p \lambda^{(i)} \vecz^{(i)} \left( \vecz^{(i)}\right)^T \,.
\label{e1_19b}
\ee
\label{e1_19}
\esube
Since $\lambda^{(1)}(1-\gamma)<0$ and $\lambda^{(i)}<0$ for $i\ge 2$ by our assumptions about
$A$ and $\gamma$, the matrix $A^{(\rm mod)}$ in (\ref{e1_19b}) is negative definite, and hence
Richardson's method (\ref{e1_11}) for it can converge.

Note that if one chooses \cite{gP}
\be
\gamma=1+1/(\lambda^{(1)}\D\tau)
\label{e1_20}
\ee
and uses the resulting
$A^{(\rm mod)}$ in Richardson's method (\ref{e1_11}), one thereby forces the projection of
the error \ $\tilde{\vecy}_n\equiv \vecy_n-\vecy$ \ on the direction of the unstable
eigenvector $\vecz^{(1)}$ to be zero at every iteration. This idea of zeroing out the
projection of the iteration error on the unstable direction lies behind the original \PM\
\cite{PelinovskyS04} and its generalization for Eq.~(\ref{e1_04}) \cite{gP}.
This is also the idea that we will use later in this paper for the development of modified CGMs.

So far we have discussed how Richardson's method for (\ref{e1_06}) with a sign indefinite
$A$ can be made to converge. The next issue is, {\em how fast} it converges. 
This is quantified by the convergence factor $R$, such that
the number of iterations required for the iteration error to decrease by a factor of $e$
is (asymptotically) inversely proportional to log$R$.  
For the optimal value of $\D\tau$, the convergence factor of Richardson's method is
(see, e.g., Sec.~5.2.2 in \cite{Axelsson_book}, or \cite{SOM}): 
\be
R=  \frac{{\rm cond}(A^{(\rm mod)})+1}{{\rm cond}(A^{(\rm mod)})-1}\,,
\label{e1_21}
\ee
where the condition number is related to the eigenvalues of $A^{(\rm mod)}$:
\be
{\rm cond}(A^{(\rm mod)})=\lambda_{\min}/\lambda_{\max}
\label{e1_22}
\ee
(recall that all the eigenvalues of $A^{(\rm mod)}$ are negative and so cond$(A^{(\rm mod)})>1$). 
For cond$(A)\gg 1$, convergence to a prescribed accuracy occurs in $O(1/\log(R))=O({\rm cond}(A))$
iterations. That is, 
the greater the condition number, the slower the convergence of the iterative method.
A common method to reduce the condition number is to use a preconditioner, which can drastically
lower $|\lambda_{\min}|$ (see, e.g., Lecture 40 in \cite{TrefethenBau_NLA}).
The preconditioned Richardson's method is
\be
\vecy_{n+1}=\vecy_n+B^{-1}(A\vecy_n-\vecb)\D\tau\,,
\label{e1_23}
\ee
where $B$ is the preconditioning matrix\footnote{
For example, the well-known Jacobi and Gauss--Seidel iterative methods reduce to
\eqref{e1_23} for appropriate choices of $B$.}. 
For the Richardson-type method
(\ref{e1_12}) with $\Lnot$ replaced by $L^{(0,\rm mod)}$, a convenient preconditioning
operator, $N$, has a form similar to that of $M$. For example, if $M$ is as in 
(\ref{e1_03}), the convenient form for $N$ is:
\be
N=c-\nabla^2, \qquad c>0\,,
\label{e1_24}
\ee
where $c$ is some constant. 

Even if the condition number is lowered by reducing $|\lambda_{\min}|$ via preconditioning,
the convergence of an iterative method may still be slow due to $|\lambda_{\max}|$ being
close to zero. In the context of the nonlinear problem (\ref{e1_04}), examplified by 
(\ref{e1_05}), this occurs when the propagation constant $\mu$ is close to the edge of
the spectral bandgap of the linear potential 
(i.e., $V_0(\cos^2x+\cos^2y)$ in (\ref{e1_05})); see, e.g., 
Fig.~4(a) in \cite{YangC06} or Fig.~\ref{fig_1} in Section 6 below. 
In \cite{ME,SOM} we proposed to accelerate the iterations by eliminating the slowest-decaying
eigenmode, i.e., the component of the error $\tu_n$ that corresponds to the eigenvalue
$\lambda_{\max}$. This effectively removes this slowest eigenmode from the spectrum of 
$L^{(\rm mod)}$ and thereby increases the magnitude of 
its maximum eigenvalue. This, in its turn, increases the convergence
rate of the method via (\ref{e1_22}) and (\ref{e1_21}). 
The slowest mode can be eliminated in exactly the same way as the unstable mode
(i.e., the mode with the positive eigenvalue $\lambda^{(1)}$) in (\ref{e1_16}).
The slowest mode can be approximated by \cite{ME}
\be
\vecz^{\rm slow} \propto \vecy_n-\vecy_{n-1}\,,
\label{e1_25}
\ee
and a preconditioned Richardson's method is then applied not to (\ref{e1_06}) but to
\be
\left( I- \gamma \vecz^{(1)} \left( \vecz^{(1)}\right)^T  -
  \gamma^{\rm slow} \vecz^{\rm slow} \left( \vecz^{\rm slow}\right)^T   \right)
\left( A\vecy - \vecb \right) = {\bf 0}\,,
\label{e1_26}
\ee
where $\gamma^{\rm slow}$ is computed similarly to (\ref{e1_20}). The explicit form
of a counterpart of
(\ref{e1_26}) for the nonlinear problem (\ref{e1_07}) will be given in Section 6.

In \cite{gP,SOM} we demonstrated that when the convergence of an iterative method is slow, 
the slowest mode elimination technique described above can considerably (by a factor of five to 
ten times) accelerate the convergence. 
However, it is well known 
(see, e.g., \cite{TrefethenBau_NLA}, p. 341 and \cite{Axelsson_book}, p. 451) 
that the fastest generic method
for a linear system (\ref{e1_06}) with a real symmetric and sign definite matrix $A$ is
the Conjugate Gradient method (CGM), whose algorithm is given by Eqs. \eqref{e2_01}
of the next section.
The convergence factor of the CGM satisfies (see, e.g., \cite{TrefethenBau_NLA}, p. 299):
\be
R \,\sim \,  \frac{ \sqrt{{\rm cond}(A)}+1}{ \sqrt{{\rm cond}(A)}-1}\,,
\label{e1_add01}
\ee
which for matrices with large condition numbers implies considerably faster convergence 
than does \eqref{e1_21}. 
Therefore, if one could modify the CGM so that it would be guaranteed to converge
when $A$ (or the linearized operator $L$) is not negative definite but has 
one positive eigenvalue, then such a modified CGM is expected to be faster than a Richardson-type
method accelerated by mode elimination. 

In this paper we present several versions of such a modified CGM for nonlinear problem \eqref{e1_07}
and its generalizations and compare them with the Richardson-type methods of \cite{gP,ITEM}
accelerated by mode elimination. We find that, as expected, the modified CGMs are the faster of these two
groups of methods. 
As we noted after Eq.~(\ref{e1_20}), 
the main idea behind all these versions of the modified CGM is to
eliminate the component of the error $\tu_n$ that is aligned along certain unstable eigenmode(s)
of the linearized iteration operator. 
Our modified CGMs are guaranteed to converge in almost all situations where a fundamental 
solitary wave with a prescribed (set of) propagation constant(s) is sought. Moreover, even
for those rare situations where our guarantee is formally void, we will show a simple and practical 
way to still make the corresponding modified CGM converge.
For the situations where a solitary wave with a prescribed (set of) power(s) is sought, 
our modified CGM is guaranteed to converge in the same parameter space where 
the Richardson-type method \cite{ITEM,vecITEM} converges.


\section{Modified CGM for solitary waves with prescribed propagation constant}
\setcounter{equation}{0}

In this section, we will first state the algorithm of the CGM for the linear Eq.~\eqref{e1_06}.
Then, we will briefly review the generalized \PM\ \cite{gP} for 
solitary waves and re-cast it in a form more
convenient for the present analysis. Finally, we will propose a modification
of the CGM for finding fundamental solitary waves with a prescribed value of the propagation
constant. All analysis in this section is done for the single-component case; its
generalization for multi-component solitary waves is presented in Section 5.

\subsection{CGM for the linear Eq.~\eqref{e1_06}}

The original CGM algorithm for Eq.~\eqref{e1_06} with a real symmetric and sign definite 
matrix $A$ and starting with an initial guess $\vecy_0$, is:
\bsube
\be
\vecr_0=A\vecy_0-\vecb, \qquad \vecd_0=\vecr_0,
\label{e2_01a}
\ee
\be
\alpha_n=-\frac{\langle \vecr_n,\,\vecd_n \rangle}{ \langle A\vecd_n,\,\vecd_n \rangle },
\label{e2_01b}
\ee
\be
\vecy_{n+1}=\vecy_n+\alpha_n\vecd_n,
\label{e2_01c}
\ee
\be
\vecr_{n+1}=A\vecy_{n+1}-\vecb,
\label{e2_01d}
\ee
\be
\beta_n= -\frac{\langle \vecr_{n+1},\,A\vecd_n \rangle}{ \langle A\vecd_n,\,\vecd_n \rangle },
\label{e2_01e}
\ee
\be
\vecd_{n+1}=\vecr_{n+1}+\beta_n\vecd_n\,.
\label{e2_01f}
\ee
\label{e2_01}
\esube
Here $\langle {\bf a},{\bf b} \rangle$ denotes the natural inner product between the
real-valued vectors ${\bf a}$ and ${\bf b}$.
Equations \eqref{e2_01a} define the initial residual vector $\vecr_0$ 
and the initial search direction $\vecd_0$. Equation \eqref{e2_01c} updates the
solution by adding to the previous solution a vector $\alpha_n\vecd_n$ 
along the search direction $\vecd_n$. The value of $\alpha_n$ is set by \eqref{e2_01b}
to guarantee the orthogonality of the new residual, defined by \eqref{e2_01d}, to
the search direction $\vecd_n$:
\be
\langle \vecr_{n+1},\,\vecd_n \rangle = 0\,.
\label{e2_02}
\ee
Finally, Eq.~\eqref{e2_01f} updates the search direction, where $\beta_n$ is set by
\eqref{e2_01e} so that
\be
\langle \vecd_{n+1},\,A\vecd_n \rangle = 0\,.
\label{e2_03}
\ee

The condition that $A$ be real symmetric is inherently required for the
derivation of algorithm \eqref{e2_01}. 
In particular, it guarantees that the orthogonality relations \eqref{e2_02} and \eqref{e2_03}
between quantities at two {\em consecutive} iterations
imply more general orthogonality relations among such quantities at {\em all} iterations
(see, e.g., \cite{Markham90} or \cite{TrefethenBau_NLA}, p. 295):
\be
\langle \vecr_{n+1},\,\vecd_j \rangle = 0\,, \quad j \le n\,,
\label{e2_add02}
\ee
\be
\langle \vecd_{n+1},\,A\vecd_j \rangle = 0\,, \quad j \le n\,.
\label{e2_add03}
\ee
The orthogonality relation \eqref{e2_add03} with a {\em sign definite} $A$ ensures that 
the CGM at each iteration produces a search direction that is linearly independent
from all previous search directions. This implies that
the error (in a certain norm) decreases with each iteration; this fact can be restated
by saying that the CGM is an optimal iterative method 
for Eq.~\eqref{e1_06} with a sign definite matrix $A$
(see, e.g., \cite{TrefethenBau_NLA}, p. 296).

The condition that $A$ be sign definite is also needed to guarantee that the denominators
of $\alpha_n$ and $\beta_n$ never vanish (or, for practical purposes, never become too close to zero). 
Thus, applying the CGM to an equation with
a sign indefinite matrix $A$ would come without the guarantee that this method 
would be optimal and that it would even converge.
However, it should be emphasized that the CGM {\em may} converge
when applied to problem (\ref{e1_06}) with a sign 
indefinite matrix $A$ (see, e.g., \cite{Markham90}, or \cite{TrefethenBau_NLA}, p. 301).
This is in stark contrast to 
Richardson's method (\ref{e1_11}), which, for a generic initial condition $\vecy_0$,
is guaranteed to diverge in such a case.

The CGM \eqref{e2_01} is straightforwardly extended to solve a nonlinear equation
\be
K^{(0)}v=0
\label{e2_add01}
\ee
whose linearized operator $K$ is self-adjoint, by replacing $A\vecy-\vecb$ with $K^{(0)}v$ 
in \eqref{e2_01a} and \eqref{e2_01d} and $A$ with $K$ in \eqref{e2_01b} and \eqref{e2_01e}.
In fact, for the CGM applied to a nonlinear problem, several choices of parameter $\beta_n$
that are equivalent in the linear approximation, will produce different versions of the
method. Two most widely used such versions are known as Fletcher--Reeves and 
Polak--Ribi\`ere (see, e.g., \cite{Shewchuk94}). In the following we will
not attempt to compare these two versions because the focus of this paper is {\em not} on
the extension of the linear algorithm \eqref{e2_01} to the nonlinear case but on the extension
of that algorithm to the case when the counterpart of matrix $A$ has one eigenvalue of the
opposite sign than the rest of the spectrum.

Let us also note that, unlike the algorithm \eqref{e2_01} for linear equations, its
extension to nonlinear problems described in the previous paragraph may fail if the
linearized operator $K$ is negative definite but has one or more eigenvalues close to zero.
This can occur if, at some iteration, the search direction $\vecd_n$ becomes aligned
primarily along the eigenmode(s) corresponding to the small eigenvalue(s). Then the denominator
in \eqref{e2_01b} will be small and hence the increment $\alpha_n\vecd_n$ of the solution
in \eqref{e2_01c}, large. If the nonlinear equation admits more than one solution for
the considered values of its parameters, which typically occurs near a bifurcation,
then a large shift, $\alpha_n\vecd_n$, of the solution may cause the iterations to converge
to a different solution than initially intended, or to diverge. Such a failure of the method
can be avoided simply by changing the initial condition $\vecy_0$. We will discuss this in more
detail in Section 6.

\subsection{Review of the generalized Petviashvili method}

In \cite{gP}, we proposed the following iterative method for finding the
fundamental solitary wave $u$ of the nonlinear problem \eqref{e1_07}:
\be
u_{n+1}-u_n = \left( N^{-1} \Lnot u_n - \gamma\,
 \frac{ \langle u_n, \Lnot u_n \rangle }{ \langle u_n, N u_n \rangle }\,u_n \right)\D \tau\,,
\label{e2_04}
\ee
where
$$
\langle f,\,g \rangle \equiv \int f(\vecx)g(\vecx) \,d\vecx
$$
and the integration is over the entire spatial domain.
The role of the $\gamma$-term is discussed at the end of this subsection.
In \eqref{e2_04}, a self-adjoint and positive definite differential operator $N$ with constant 
coefficients is chosen so that $u$ is an approximate eigenvalue of $N^{-1}L$ corresponding
to its largest eigenvalue $\lambda^{(1)}$:
\be
N^{-1}Lu \approx \lambda^{(1)}u\,.
\label{e2_05}
\ee
The Sylvester law of inertia (see, e.g., \cite{HornJohnson_book})
guarantees that the signs of the eigenvalues of $N^{-1}L$ are the same as the signs of the
respective eigenvalues of $L$. Then, 
according to our definition of a fundamental solitary wave at the end of Introduction,
$\lambda^{(1)}$ may be positive and all the other eigenvalues of $N^{-1}L$ are negative.

The explicit form of $N$ is usually taken to be similar to that of $M$. 
For example, when $M$ is given by (\ref{e1_03}), $N$ has
the form (\ref{e1_24}) where the constant $c$ can be computed algorithmically 
from $u_n$ \cite{gP}.
(For the equation (\ref{e1_01}) with power-law nonlinearity, $N=M$ and the equality
in (\ref{e2_05}) is exact \cite{PelinovskyS04}.) The constant 
$\gamma$ in \eqref{e2_04}
is given by \eqref{e1_20} and $\lambda^{(1)}\equiv\lambda^{(1)}_n$ is found from
\be
\lambda^{(1)}_n= \langle u_n, \,L u_n \rangle / \langle u_n, \,N u_n \rangle\,.
\label{e2_06}
\ee
In practice, $N$, $\gamma$, and $\lambda^{(1)}$ are computed until the iteration error 
reaches some predefined
tolerance, after which their last-computed values are used for all subsequent iterations.
Since $N$ is a differential operator with constant coefficients, it has a simple
representation in the Fourier space. Therefore, quantities like $N^{-1}\Lnot u_n$,
\ $Nu_n$, etc. are easily computed using the direct and inverse Fast Fourier Transforms,
which are available as built-in commands in all major computing software.

In what follows we will frequently refer to the linearized form of \eqref{e2_04}, which is:
\be
\tu_{n+1}-\tu_n = \left( N^{-1} L \tu_n - \gamma\,
 \frac{ \langle u, \,L\tu_n \rangle }{ \langle u, N u \rangle }\,u \right)\D \tau\,.
\label{e2_07}
\ee
Note that the operator $N^{-1}L$ in the leading term of \eqref{e2_07} is not self-adjoint.
Therefore, we introduce a commonly employed
change of variables to rewrite that equation in a form involving a
self-adjoint operator. Namely, we denote
\bsube
\be
v=N^{1/2}u, \quad \tv_n=N^{1/2}\tu_n, \quad K=N^{-1/2}L\,N^{-1/2}\,,
\label{e2_08a}
\ee
\be
\Knot v_n \equiv N^{-1/2}\Lnot u_n\,.
\label{e2_08b}
\ee
\label{e2_08}
\esube
Note that operator $K$ is self-adjoint, since $L$ is self-adjoint and $N$ is self-adjoint
and positive definite.
Let us stress that $K$ and $v$ have been introduced only for the convenience of 
carrying out the subsequent analysis with self-adjoint operators. Once the derivation
of the algorithm in this framework is complete, we will recast it in terms of the
original variable $u_n$ and operator $L$. 

With the notations \eqref{e2_08}, 
Eqs. \eqref{e2_05}, \eqref{e2_04}, and \eqref{e2_07} are rewritten as:
\be
Kv\approx \lambda^{(1)} v\,,
\label{e2_09}
\ee
\be
v_{n+1}-v_n = \left(  \Knot v_n - \gamma\,
 \frac{ \langle v_n, \Knot v_n \rangle }{ \langle v_n, v_n \rangle }\,v_n \right)\D \tau\,,
\label{e2_10}
\ee
\be
\tv_{n+1}-\tv_n = \left( K \tv_n - \gamma\,
 \frac{ \langle v, \,K\tv_n \rangle }{ \langle v, v \rangle }\,v \right)\D \tau\,.
\label{e2_11}
\ee
Equations \eqref{e2_10} and \eqref{e2_11} are the nonlinear and linearized counterparts of
the linear problem \eqref{e1_16} with a symmetric matrix $A$. The $\gamma$-terms in these
equations are needed to
eliminate the component of the error $\tv_{n+1}$ along the unstable ``direction"
$v$; see \eqref{e2_09} and the text after \eqref{e2_05}. 
If the equality in \eqref{e2_09} (and \eqref{e2_05}) is exact, which occurs only for
equations with power-law nonlinearity, \eqref{e1_01}, this elimination is complete.
However, for the majority of equations \eqref{e1_04} with a more complicated nonlinear
term, $v$ is not an exact eigenmode of $K$ and hence the $\gamma$-terms in 
\eqref{e2_10} and \eqref{e2_11} strongly suppress, but do not completely eliminate, 
the component along the unstable eigenmode. Nonetheless, this strong suppression is
sufficient to turn the unstable eigenmode into a stable one. 

In the remainder of this section, we will continue using Eq.~\eqref{e2_11} with the 
self-adjoint operator $K$ for all derivations, and only at the final steps of those
derivations will convert to the original variable $u_n$.

\subsection{A modified CGM}

As we noted at the end of Sec. 3.1, 
the straightforward nonlinear generalization of the CGM applied to Eq.~\eqref{e2_add01}
whose linearized operator $K$ is self-adjoint but not sign definite, may diverge. 
For equations that we consider in Section 6 below, this method indeed diverges.
To modify the method so that it could converge, one can use the idea stated before
Eq.~\eqref{e1_14}. Namely, replace \eqref{e2_add01} with an equivalent equation \ 
$K^{(0, \rm mod)}v=0$ such that the corresponding linearized operator $K^{(\rm mod)}$ is
self-adjoint and sign definite. In Appendix 1 we
show that the expression in \eqref{e2_12a} below, which mimics that on the right-hand side of
\eqref{e2_10}, comes close to satisfying these properties: (i) it is equivalent to 
$\Knot v$, (ii) its linearized operator $K^{(\rm mod)}$ is approximately 
self-adjoint, and  (iii) its eigenvalues that are not too close to zero are all negative. 
Statements (ii) and (iii) are approximate\footnote{
However, as we point out in Appendix 1, in all the examples that we tried and
some of which are reported in Section 6, statement (iii) holds in a more definite sense, i.e.:
{\em all} eigenvalues of $K^{(\rm mod)}$ are negative.
}
rather than exact because so is relation \eqref{e2_09}. 
We will discuss the implications of the approximate character of statement (ii) after
we present the algorithm of the modified CGM.

Thus, denoting 
\bsube
\be
K^{(0, \rm mod)}v = \Knot v - \Gamma\,
 \frac{ \langle v, \Knot v \rangle }{ \langle v, v \rangle }\,v\,,
\label{e2_12a}
\ee
\be
\Gamma = 1 + \frac1{\lambda^{(1)}}\,,
\label{e2_12b}
\ee
\be
K^{(\rm mod)}\tv = K \tv - \Gamma\,
 \frac{ \langle v, \,K\tv \rangle }{ \langle v, v \rangle }\,v\,,
\label{e2_12c}
\ee
\label{e2_12}
\esube
we propose the following modified CGM for finding the fundamental solitary 
wave of \eqref{e2_add01}.
It mimics algorithm \eqref{e2_01} with the following modifications:
$A\vecy-\vecb$ in \eqref{e2_01a} and \eqref{e2_01d} is replaced with $K^{(0, \rm mod)}v$, 
$A$ in \eqref{e2_01b} and \eqref{e2_01e} is replaced with $K^{(\rm mod)}$,
and vectors $\vecr$ and $\vecd$ are renamed $\bar{r}$ and $\bar{d}$. Upon the
change of variables 
\be
\bar{r}_n=N^{1/2}r_n, \qquad \bar{d}_n=N^{1/2}d_n,
\label{e2_13}
\ee
and \eqref{e2_08}, this modified CGM is:
\bsube
\be
r_0= N^{-1}L^{(0,\rm mod)}u_0, \qquad d_0=r_0,
\label{e2_14a}
\ee
\be
\alpha_n=-
\frac{\langle Nr_n,\,d_n \rangle}{ \langle L^{(\rm mod)}d_n,\,d_n \rangle },
\label{e2_14b}
\ee
\be
u_{n+1}=u_n+\alpha_n d_n,
\label{e2_14c}
\ee
\be
r_{n+1}= N^{-1}L^{(0,\rm mod)}u_{n+1}
\label{e2_14d}
\ee
\be
\beta_n= -
 \frac{\langle r_{n+1},\,L^{(\rm mod)}d_n \rangle}{ \langle L^{(\rm mod)}d_n,\,d_n \rangle },
\label{e2_14e}
\ee
\be
d_{n+1}=r_{n+1}+\beta_n d_n\,.
\label{e2_14f}
\ee
\label{e2_14}
\esube
where
\bsube
\be
L^{(0, \rm mod)}u = \Lnot u - \Gamma\,
 \frac{ \langle u, \Lnot u \rangle }{ \langle u, Nu \rangle }\,Nu\,,
\label{e2_15a}
\ee
\be
L^{(\rm mod)}d = L d - \Gamma\,
 \frac{ \langle u,\, L d \rangle }{ \langle u, Nu \rangle }\,Nu\,.
\label{e2_15b}
\ee
\label{e2_15}
\esube

In a practical implementation of algorithm \eqref{e2_14}, iterations should be carried out 
with the generalized \PM\ \eqref{e2_04} until the error reaches some predefined tolerance
(usually, between 1 and 10\%),
so that the computed parameters of operator $N$ and the constants $\lambda^{(1)}$
and $\Gamma$ may be considered as known with the corresponding accuracy. After that,
the iterations should be carried out with the modified CGM \eqref{e2_14}, which is expected
to converge considerably faster than the generalized \PM.

Operator $K^{(\rm mod)}$ in \eqref{e2_12c}, and hence the quadratic form
$\langle \bar{d}_n,\,K^{(\rm mod)}\bar{d}_n\rangle$ in the counterparts of 
(\ref{e2_14}b,e), would be guaranteed to be negative definite 
if in relation \eqref{e2_09} (or, equivalently, in \eqref{e2_05}), the equality held exactly.
However, this takes place only for equations with power-law nonlinearity, \eqref{e1_01}. 
Therefore, for some nonlinear wave equations, 
$\langle \bar{d}_n,\,K^{(\rm mod)}\bar{d}_n\rangle$ could be sign indefinite and, importantly,
arbitrarily close to zero. In such situations 
algorithm \eqref{e2_14} could fail, and we will now point out when this should be expected.
As we explain in Appendix 1, the slight non-self-adjointness of
$K^{(\rm mod)}$ can cause the quadratic form 
to become too close to zero when some of the negative eigenvalues of
the linearized operator $L$ of \eqref{e1_07} are close to zero.
In single-component equations, this typically occurs when the solitary wave
bifurcates from the edge of the continuous spectrum, while for multi-component equations
this can also happen when, say, an asymmetric wave bifurcates from one with a symmetry
(see, e.g., \cite{DCDP_stability} and references therein). 

Our numerical experiments confirm that algorithm \eqref{e2_14} 
converges whenever the eigenvalues of $L$ are sufficiently far from 
zero\footnote{
Note that if the solitary wave is translationally invariant, $L$ has a zero eigenvalue
corresponding to the respective translational eigenmode. 
As we show in Appendix 1, such a mode presents no problem
for convergence of the iterations since it would simply slightly shift the solitary wave.
Therefore, in the following we ignore the possible presence of such zero eigenvalues of $L$.
},
and also that is may diverge in the opposite situation, as described above. 
Recall, from the last paragraph of Section 3.1, that in exactly the same situation,
a CGM for a {\em nonlinear} problem can fail even if the quadratic form 
$\langle \bar{d}_n,\,K^{(\rm mod)}\bar{d}_n\rangle$ were negative definite. 
Thus, there are two different reasons that can cause the modified
CGM to fail, but, fortunately for the applications of the method, they both
can do so in the same situation, namely, when $L$ has near-zero eigenvalues.
Moreover, they both require that at some iteration, the search direction become primarily ``aligned"
with the eigenmode of $L$ corresponding to a small eigenvalue. 
Therefore, a simple and practical way to avoid such divergence is to change the initial condition 
in a certain manner. In Section 6 we show that this way does indeed work.

Before concluding this section, let us note that one could develop a modified version of the CGM 
based on a different idea than the elimination of the unstable eigenmode from the 
underlying operator, as in \eqref{e2_12}. Instead, one can employ the original operators
$\Knot$ and $K$ but force all search directions
$\bar{d}_n$ to be orthogonal to the unstable eigenmode of $K$ 
approximated by the exact solitary wave $v$.
Since \eqref{e2_09} holds approximately, 
the search directions can be made only approximately orthogonal to the true unstable eigenmode
of $K$. One can show that this causes this other modified CGM to be prone to failure
under the same circumstances as the modified CGM \eqref{e2_14}, i.e. when $L$ has
eigenvalues that are close to zero. Our numerical experiments show that the
simple trick that can help algorithm \eqref{e2_14} overcome that failure, is not as effective
for the other modified CGM. Moreover, the coding 
of the latter method is slightly more complicated than that of
\eqref{e2_14}. For these reasons, we only present that alternative modified CGM in Appendix 2,
but do not discuss it further in this paper.


\section{Modified CGM for solitary waves with prescribed power}
\setcounter{equation}{0}

We first review the preconditioned imaginary-time evolution method (ITEM) \cite{ITEM},
which is a Richardson-type iterative
method for finding solitary waves with a specified value of power,
and then present a modified CGM that converges under the same conditions as the ITEM,
but faster.

\subsection{Review of the ITEM}

When one seeks a solitary wave with a prescribed value of power \eqref{e1_27},
problem \eqref{e1_07} should be redefined because the propagation constant $\mu$,
which enters $\Lnot$ via operator $M$ (see \eqref{e1_03}), is no longer known
exactly. Instead, it is estimated at each iteration, as shown below. Thus, we rewrite 
\eqref{e1_04} as 
\bsube
\be
\Lnnot u - \mu u=0,
\label{e3_01a}
\ee
where
\be
\Lnnot u = \nabla^2 u + F(u,\vecx), \qquad 
\mu= \frac{ \langle f(u), \Lnnot u \rangle}{ \langle f(u), u \rangle } \,,
\label{e3_01b}
\ee
\label{e3_01}
\esube
and $f(u)$ is any function of $u$; its choice will be specified shortly.

The preconditioned ITEM whose convergence conditions are found in \cite{ITEM} is:
\bsube
\be
\mu_n= \frac{ \langle N^{-1}u, \Lnnot u \rangle}{ \langle N^{-1}u, u \rangle } \,,
\label{e3_02a}
\ee
\be
\hat{u}_{n+1}=u_n+N^{-1} \left( \Lnnot u_n - \mu_n u_n \right) \D\tau\,,
\label{e3_02b}
\ee
\be
u_{n+1}=\hat{u}_{n+1} \sqrt{ \frac{P}{\langle \hat{u}_{n+1}, \hat{u}_{n+1} \rangle } } \,.
\label{e3_02c}
\ee
\label{e3_02}
\esube
Here $P$ is the prescribed value of the power and $N$ is a preconditioning operator
of the form \eqref{e1_24} where now $c$ is an {\em arbitrary} (unlike in Section 3)
positive number.

In what follows we will need a few facts about the linearized method \eqref{e3_02} 
\cite{ITEM}.
First, the condition that \ $ \langle u_n, u_n \rangle = P={\rm const}$ \ entails
the orthogonality between the error and the exact solution:
\be
\langle \tu_n, u \rangle = O(\tu_n^2)\,.
\label{e3_03}
\ee
Next, the linearized form of the operator in parentheses in \eqref{e3_02b} is:
\be
\Lcal \tu_n \equiv  L\tu_n - 
  \frac{ \langle N^{-1}u, L \tu_n \rangle}{ \langle N^{-1}u, u \rangle } u\,.
\label{e3_04}
\ee
(Note that operator $L$ involves the propagation constant $\mu$.
Even though $\mu$ is not specified when we seek the solitary wave, we can still
use it in analyses of iterative methods.)
Using the last two equations, one can straightforwardly
show that the last line, \eqref{e3_02c}, of
the ITEM does not change the linearization of the preceding line. Equation \eqref{e3_02c}
is needed to ensure that the power of the solitary wave equals $P$ {\em exactly} rather
than in the linear approximation. 

The ITEM \eqref{e3_02} converges when its linearized operator, $\Lcal$, has only
negative eigenvalues (see the footnote at the end of Section 3). 
Loosely speaking (see \cite{ITEM} for a more precise statement), 
for a wide subclass of equations \eqref{e1_04},
this occurs when operator $L$ has none or one positive eigenvalues. Moreover,
in the latter case, the following condition has to hold:
\be
dP/d\mu > 0\,.
\label{e3_05}
\ee

Below we will use the change of variables \eqref{e2_08a} and similar notations:
\be
\Knnot v_n \equiv N^{-1/2} \Lnnot u_n\,, \qquad
\Kcalnot v_N \equiv \Knnot v_n - \mu_n N^{-1} v_n\,,
\label{e3_06}
\ee
where $\mu_n$ is given by \eqref{e3_02a}. The linearization of $\Kcalnot v_n$ is
an operator
\be
\Kcal \tv_n \equiv K\tv_n - 
 \frac{ \langle N^{-1}v,K\tv_n \rangle }{ \langle N^{-1}v, N^{-1}v \rangle } N^{-1}v\,.
\label{e3_07}
\ee
When the convergence condition of the ITEM
\eqref{e3_02}, stated near \eqref{e3_05}, hold, operator $\Kcal$ 
is \cite{ITEM} self-adjoint and negative definite 
{\em on the space of functions satisfying the exact form of the
orthogonality relation \eqref{e3_03}}:
\be
\langle \tv_n, N^{-1}v \rangle = 0\,.
\label{e3_08}
\ee
%


\subsection{A modified CGM}

The idea of this modified CGM is to ensure that both the iteration error $\tv_n$ and the
search direction be orthogonal, in the sense of \eqref{e3_08}, to the exact solution $v$.
Then, according to the last sentence in the previous subsection, operator $\Kcal$ is
guaranteed to be negative definite on this space of functions, and hence the following
modified CGM is guaranteed to converge under the same conditions as the ITEM \eqref{e3_02}:
\bsube
\be
\bar{r}_0= \Kcalnot v_0, \qquad 
\bar{d}_0=\bar{r}_0 - \frac1{P}\langle N^{-1} v_0, \bar{r}_0 \rangle \, v_0,
\label{e3_09a}
\ee
\be
\alpha_n=-
\frac{ \dst \langle \bar{r}_n,\,\bar{d}_n \rangle }
{ \langle \bar{d}_n,\,\Kcal \bar{d}_n \rangle }\,,
\label{e3_09b}
\ee
\be
\hat{v}_{n+1}=v_n+\alpha_n \bar{d}_n\,, \qquad
v_{n+1}=\hat{v}_{n+1} \sqrt{\frac{P}{ \langle \hat{v}_{n+1}, N^{-1} \hat{v}_{n+1}\rangle }}\,,
\label{e3_09c}
\ee
\be
\bar{r}_{n+1}= \Kcalnot v_{n+1}
\label{e3_09d}
\ee
\be
\beta_n= -
 \frac{ \dst \langle \bar{r}_{n+1},\,\Kcal \bar{d}_n \rangle - 
   \frac1{P} 
   \langle v_{n+1}, \Kcal \bar{d}_n \rangle \langle N^{-1}v_{n+1}, \bar{r}_{n+1} \rangle  }
{ \langle \bar{d}_n,\,\Kcal\bar{d}_n \rangle },
\label{e3_09e}
\ee
\be
\bar{d}_{n+1}= \bar{r}_{n+1}+\beta_n \bar{d}_n - 
 \frac1P \langle N^{-1}v_{n+1}, \,\bar{r}_{n+1}+\beta_n \bar{d}_n \rangle \, v_{n+1} \,.
\label{e3_09f}
\ee
\label{e3_09}
\esube
The definitions of $\bar{d}_0$ in \eqref{e3_09a} and $\bar{d}_{n+1}$ in \eqref{e3_09f}
ensure that
\be
\langle \bar{d}_n, N^{-1} v_n \rangle = 0
\label{e3_10}
\ee
at every iteration. 
Here we use the fact that our iterative solution has a fixed value of power:
\be
\langle v_n, N^{-1} v_n \rangle = P
\label{e3_11}
\ee
for all $n$.
Since the error $\tv_n$ at the $n$th iteration satisfies
the orthogonality condition \eqref{e3_08}, the first of Eqs. \eqref{e3_09c} and 
Eq. \eqref{e3_10} ensure that the error $\tv_{n+1}$ at the next iteration also
satisfies \eqref{e3_08}. The second of Eqs. \eqref{e3_09c} does not change this
relation; it only forces \eqref{e3_11} to be satisfied
exactly rather than in the linear approximation. 
Equations (\ref{e3_09}b--d) entail the counterpart of \eqref{e2_02}:
\be
\langle \bar{r}_{n+1},\,\bar{d}_n \rangle = O(\tv_n^3)\,,
\label{e3_12}
\ee
and Eqs.~(\ref{e3_09}e,f) entail a counterpart of \eqref{e2_03}:
\be
\langle \bar{d}_{n+1},\,\Kcal\bar{d}_n \rangle = O(\tv_n^3)\,.
\label{e3_13}
\ee
We demonstrate these relations in Appendix 1.

Let us note that condition \eqref{e3_10}, in addition to ensuring \eqref{e3_08} at each 
iteration, playes one other important role in this algorithm. Namely, it ensures that 
in the inner product on the left-hand side of \eqref{e3_13}, $\Kcal$ is a self-adjoint
operator: see the sentence before \eqref{e3_08}. 
This property of $\Kcal$ entails the counterpart of \eqref{e2_add03}, which together with 
negative definiteness of $\Kcal$ implies that all search directions
$\bar{d}_n$ are linearly independent and the error decreases as the iterations proceed.
In Section 3.1 we noted that this means that the CGM is an optimal iterative method 
(see, e.g., \cite{TrefethenBau_NLA}, p. 296, and \cite{Shewchuk94}). 

Thus, the modified CGM \eqref{e3_09} is guaranteed to converge under the same conditions
as its Richardson-type counterpart \eqref{e3_02}. This, mathematically, is a more rigorous
result than what we obtained for the modified CGM \eqref{e2_14}, which is guaranteed to converge 
in a slightly narrower parameter space than its Richardson-type counterpart \eqref{e2_04}.

Finally, we present algorithm \eqref{e3_09} in the original variables:
\bsube
\be
r_0= N^{-1} \Lcalnot u_0, \qquad 
d_0=r_0 - \frac1{P} \langle u_0, r_0 \rangle\,u_0\,,
\label{e3_14a}
\ee
\be
\alpha_n=-
\frac{ \langle r_n,\,N d_n \rangle }
{ \langle d_n,\,\Lcal d_n \rangle }\,,
\label{e3_14b}
\ee
\be
\hat{u}_{n+1}=u_n+\alpha_n d_n\,, \qquad
u_{n+1}=\hat{u}_{n+1} \sqrt{\frac{P}{ \langle \hat{u}_{n+1}, \hat{u}_{n+1} \rangle }}\,,
\label{e3_14c}
\ee
\be
r_{n+1}= N^{-1} \Lcalnot u_{n+1}
\label{e3_14d}
\ee
\be
\beta_n= -
 \frac{ \dst \langle r_{n+1},\,\Lcal d_n \rangle - 
   \frac1{P} 
   \langle u_{n+1}, \Lcal d_n \rangle \langle u_{n+1}, r_{n+1}\rangle  }
{ \langle d_n,\,\Lcal d_n \rangle },
\label{e3_14e}
\ee
\be
d_{n+1}= r_{n+1}+\beta_n d_n - 
 \frac1{P} \langle u_{n+1} , \, r_{n+1}+\beta_n d_n \rangle \, u_{n+1} \,.
\label{e3_14f}
\ee
\label{e3_14}
\esube
Here $\Lcal$, which is the linearization of 
\be
\Lcalnot u_n \equiv \Lnnot u_n - 
 \frac{ \langle N^{-1}u_n, \Lnnot u_n \rangle}{ \langle N^{-1}u_n, u_n \rangle } \, u_n\,,
\label{e3_15}
\ee
is given in \eqref{e3_04}.


\section{Modified CGMs for multi-component solitary waves}
\setcounter{equation}{0}

Here we present the multi-component versions of the modified CGMs \eqref{e2_14} and
\eqref{e3_14} for fundamental solitary waves with prescribed values of,
respectively, propagation constants and quadratic conserved quantities that are
the counterparts of power \eqref{e1_27}. We will not give derivations of these methods
because they are similar to those found in Sections 3 and 4. 

\medskip

We begin with stating the generalized \PM\ from \cite{gP} for finding a multi-component
solitary wave \ ${\bf u}=[ u^{(1)},\,\ldots\,, u^{(S)} ]^T$ of the equation
\be
{\bf \Lnot}{\bf u} = {\bf 0}
\label{e4_01}
\ee
whose linearized operator is ${\bf L}$:
\be
{\bf u}_{n+1}={\bf u}_n + \left[ {\bf N}^{-1} {\bf \Lnot}{\bf u}_n - 
 \sum_{k=1}^S \gamma^{(k)} 
 \frac{ \left\langle {\bf e}^{(k)}_n,\, {\bf \Lnot}{\bf u}_n \right\rangle }
      { \left\langle {\bf e}^{(k)}_n,\,{\bf N} {\bf e}^{(k)}_n \right\rangle}\,
      {\bf e}^{(k)}_n \, \right] \D\tau\,.
\label{e4_02}
\ee
Here ${\bf N}={\rm diag}\left( N^{(1)},\ldots\,, N^{(S)} \right)$ where each $N^{(k)}$
has a form similar to the linear differential part of the $k$th component of ${\bf \Lnot}$;
e.g., for a system of nonlinear Schr\"odinger-type equations, 
\be
N^{(k)}=c^{(k)}-b^{(k)}\nabla^2,
\label{e4_03}
\ee
and
$c^{(k)}$, $b^{(k)}$ are computed by formulas found in \cite{gP}. Furthermore,
${\bf e}^{(k)}$ are the approximate eigenvectors of ${\bf N}^{-1}{\bf L}$:
\be
{\bf L}{\bf e}^{(k)} \approx \lambda^{(k)} {\bf N} {\bf e}^{(k)}\,, 
\label{e4_04}
\ee
which are mutually orthogonal with weight ${\bf N}$:
\be
\langle {\bf e}^{(k)}, {\bf N} {\bf e}^{(m)} \rangle = 0 \qquad \mbox{for $k\neq m$.}
\label{e4_05}
\ee
Note that \eqref{e4_04} is the multi-component counterpart of \eqref{e2_05}.
The form of ${\bf e}^{(k)}$'s is related to the solution of \eqref{e4_01} by
\be
{\bf e}^{(1)}_n={\bf u}_n, \quad  
{\bf e}^{(k)}= \left( \rho^{(k,1)} u^{(1)}, \ldots\,, \rho^{(k,S)} u^{(S)} \right)^T\,,
\label{e4_06}
\ee
where $\rho^{(k,k)}=1$ and the rest of $\rho^{(k,j)}$ are found as explained in \cite{gP}.
Finally, the constants $\lambda^{(k)}$ and 
$\gamma^{(k)}$ in \eqref{e4_02} are found similarly to \eqref{e2_06} and \eqref{e1_20}:
\be 
\lambda^{(k)} = \langle {\bf e}^{(k)}, {\bf L} {\bf e}^{(k)} \rangle /
                \langle {\bf e}^{(k)}, {\bf N} {\bf e}^{(k)} \rangle\,, 
\qquad
\gamma^{(k)}=1+1/(\lambda^{(k)}\D\tau)\,.
\label{e4_07}
\ee

The multi-component version of the modified CGM of Section 3 looks as \eqref{e2_14},
with the scalar functions $u_n$, $r_n$, $d_n$ being replaced by the $S$-dimensional vectors
${\bf u}_n$, $\vecr_n$, $\vecd_n$, operator $N$ being replaced by its matrix counterpart
given before Eq.~\eqref{e4_03}, and the operators ${\bf L^{(0,\rm mod)}}$ and ${\bf L^{(\rm mod)}}$
being given by expressions generalizing \eqref{e2_15}:
\bsube
\be
{\bf L^{(0,\rm mod)} u}_n \,=\, {\bf \Lnot}{\bf u}_n - 
 \sum_{k=1}^S \Gamma^{(k)} 
 \frac{ \left\langle {\bf e}^{(k)}_n,\, {\bf \Lnot}{\bf u}_n \right\rangle }
      { \left\langle {\bf e}^{(k)}_n,\,{\bf N} {\bf e}^{(k)}_n \right\rangle}\,
      {\bf N\, e}^{(k)}_n \,,
\label{e4_08a}
\ee
\be
\Gamma^{(k)} = 1 +\frac1{\lambda^{(k)}}\,,
\label{e4_08b}
\ee
\be
{\bf L^{(\rm mod)} d}_n \,=\, {\bf L}{\bf d}_n - 
 \sum_{k=1}^S \Gamma^{(k)} 
 \frac{ \left\langle {\bf e}^{(k)}_n,\, {\bf L}{\bf d}_n \right\rangle }
      { \left\langle {\bf e}^{(k)}_n,\,{\bf N} {\bf e}^{(k)}_n \right\rangle}\,
      {\bf N\, e}^{(k)}_n \,.
\label{e4_08c}
\ee
\label{e4_08}
\esube
A sample code of this algorithm for a two-component system considered in Section 6
is presented in Appendix 3.

Let us note that the purpose of the terms under the sum in \eqref{e4_08} is to 
eliminate the eigenmodes ${\bf e}^{(k)}$ of ${\bf L}$ whose eigenvalues $\lambda^{(k)}$
could be positive. By our definition of the fundamental solitary wave, stated at the
end of Introduction, eliminating $S$ such eigenmodes should be sufficient to turn the
positive eigenvalues of ${\bf L}$ into negative eigenvalues of ${\bf L^{(\rm mod)}}$.
As explained in Section 3, due to the approximate character of relations \eqref{e4_04}, this
unstable mode elimination still may not always
guarantee convergence of the multi-component
modified CGM \eqref{e2_14}, \eqref{e4_08} when ${\bf L}$ has 
small negative eigenvalues.
However, a simple approach based on shifting the initial condition, 
as discussed in Section 6,  overcomes this divergence.

\bigskip

We now turn to the case where a set of quadratic conserved quantities related to the
solution components' powers are prescribed. We need to introduce a number of new notations.
Let us denote the $s$-component 
vector of these quantities by $\vecQ$, so that its $k$th component is
\be
Q^{(k)}= \sum_{l=1}^S q^{(kl)} P^{(l)}, \quad P^{(l)}=\langle u^{(l)},u^{(l)} \rangle,
\qquad k=1,\ldots\,, \,s \le S\,, \quad l=1,\ldots\,,\, S.
\label{e4_09}
\ee
Note that the number of these conserved quantities can be less than the number of
the components of the solitary wave: $s\le S$. To emphasize this fact, we use a different
vector notation for $\vecQ$ than for ${\bf u}$. The number of the components of $\vecQ$
equals the number of the propagation constants that one can prescribe independently.
One example of this situation is the system of three waves interacting via quadratic
nonlinearity \cite{3w_78} (see also \cite{SOM}). Another example, which we discuss in
more detail below to make our notations clear, is the system describing
the evolution of pulses in a two-core nonlinear directional coupler \cite{DCDP_stability}:
\be
\ba{l}
\dst
iU^{(k)}_t+U^{(k)}_{xx}+ \left( |U^{(k)}|^2+\kappa |U^{(k+2)}|^2 \right) U^{(k)} + U^{(3-k)}= 0 
\vspace{0.2cm} \\
iU^{(k+2)}_t+U^{(k+2)}_{xx}+ \left( |U^{(k+2)}|^2+\kappa |U^{(k)}|^2 \right) U^{(k+2)} + 
 U^{(5-k)}= 0
\ea
\qquad k=1,2\,,
\label{e4_10}
\ee
where $U^{(k)}$ and $U^{(k+2)}$ are the orthogonal 
polarization components of the pulse in the $k$th core.
Upon the substitution 
\be
\left( \ba{c} U^{(1)}(x,t) \\ U^{(2)}(x,t) \\ U^{(3)}(x,t) \\ U^{(4)}(x,t) \ea \right)  = 
\left( \ba{cc} u^{(1)}(x) & 0 \\ u^{(2)}(x) & 0 \\ 0 & u^{(3)}(x) \\ 0 & u^{(4)}(x) \ea \right)
\left( \ba{c} \dst e^{i\mu^{(1)}t} \\ \dst e^{i\mu^{(2)}t} \ea \right)\,,
\label{e4_11}
\ee
where $u^{(k)}$ can be chosen to be real-valued, 
system \eqref{e4_10} reduces to:
\be
\left( \ba{l} u^{(1)}_{xx}+ \left( (u^{(1)})^2+\kappa (u^{(3)})^2 \right) u^{(1)} + u^{(2)} 
              \vspace{0.1cm} \\
              u^{(2)}_{xx}+ \left( (u^{(2)})^2+\kappa (u^{(4)})^2 \right) u^{(2)} + u^{(1)} 
              \vspace{0.1cm} \\
              u^{(3)}_{xx}+ \left( (u^{(3)})^2+\kappa (u^{(1)})^2 \right) u^{(3)} + u^{(4)} 
              \vspace{0.1cm} \\
              u^{(4)}_{xx}+ \left( (u^{(4)})^2+\kappa (u^{(2)})^2 \right) u^{(4)} + u^{(3)} 
              \ea \right)  \; - \; 
   \left( \ba{cc} u^{(1)} & 0 \\ u^{(2)} & 0 \\ 0 & u^{(3)} \\ 0 & u^{(4)} \ea \right)\,\vecmu
    \;=\;  \left( \ba{c} 0 \\ 0 \\ 0 \\ 0 \ea \right)   \,, 
   \label{e4_12}
   \ee
where $\vecmu= \left( \mu^{(1)},\, \mu^{(2)} \right)^T$. Thus, in this example, $S=4$, $s=2$,
and the vector of quadratic
conserved quantities (verified from the time-dependent equations \eqref{e4_10}) is
\be
\vecQ = \left( \ba{c} P^{(1)}+P^{(2)} \\ P^{(3)}+P^{(4)} \ea \right) \equiv
 \left( \ba{cccc} 1 & 1 & 0 & 0 \\ 0 & 0 & 1 & 1 \ea \right) \
 \left( \ba{c} P^{(1)} \\ P^{(2)} \\ P^{(3)} \\ P^{(4)} \ea \right)\,.
\label{e4_13}
\ee

Generalizing the above example, one can see that the counterpart of Eq.~\eqref{e4_01}
when $\vecQ$ rather than $\vecmu$ is prescribed, is the following extension of Eq.~\eqref{e3_01a}:
\bsube
\be
{\bf \Lnnot}{\bf u} - \Ucal \, \langle {\bf N}^{-1} \Ucal,\, \Ucal \rangle^{-1}\,
  \langle {\bf N}^{-1} \Ucal, \, {\bf \Lnnot} {\bf u} \rangle \,=\, {\bf 0}\,,
\label{e4_14a}
\ee
\be
\Ucal \equiv \frac{\delta \vecQ}{\delta {\bf u}}\,.
\label{e4_14b}
\ee
\label{e4_14}
\esube
For example, in \eqref{e4_12}, ${\bf \Lnnot u}$ is the first term (the $4\times 1$ vector), and 
$\Ucal$ is the first factor of the second term (the $4\times 2$ matrix) on the left-hand side.

For the convenience of subsequent notations, we will assume that 
the matrix $\big( q^{(kl)} \big)$ in \eqref{e4_09} is put into the reduced echelon form
(see any textbook on undergraduate linear algebra) and, in addition, its columns are rearranged
so that
\be
q^{(kl)} \,=\, \left\{ \ba{ll} 0, & l<k \\ 1, & l=k\,. \ea \right. 
\label{e4_15}
\ee
For example, for the $2\times 4$ matrix in \eqref{e4_13} this would imply switching the second
and third columns. Then for the powers of the solitary wave components whose indices equal the
indices of the pivot columns of $\big( q^{(kl)} \big)$, one has from \eqref{e4_09}:
\be
P^{(k)} = Q^{(k)}- \sum_{l=k+1}^S q^{(kl)} P^{(l)}\,, \qquad k=1,\ldots\,,\, s\le S\,.
\label{e4_16}
\ee
We will use this fact in the next paragraph.

With the convention \eqref{e4_15}, the multi-component version of the ITEM is:
\bsube
\be
\hat{\bf u}_{n+1}={\bf u}_n + {\bf N}^{-1} \left( {\bf \Lnnot u}_n - 
 \Ucal_n \, \langle {\bf N}^{-1} \Ucal_n,\, \Ucal_n \rangle^{-1}\,
  \langle {\bf N}^{-1} \Ucal_n, \,  {\bf \Lnnot} {\bf u}_n \rangle  \right) \D\tau\,,
\label{e4_17a}
\ee
\be
u^{(k)}_{n+1} = \hat{u}^{(k)}_{n+1} 
\sqrt{ \frac{ Q^{(k)}- \sum_{l=k+1}^S q^{(kl)} \hat{P}^{(l)}_{n+1} }{ \hat{P}^{(k)}_{n+1} } }\;, 
\qquad    k=1,\ldots\,,\, s\le S\,,
\label{e4_17b}
\ee
\label{e4_17}
\esube
where
$$
\hat{P}^{(k)}_{n+1} \equiv \langle \hat{u}^{(k)}_{n+1},\, \hat{u}^{(k)}_{n+1} \rangle\,,
\qquad  k=1,\ldots\,,\,S\,.
$$
In analogy to the algorithm \eqref{e3_02} for single-component equations, 
Eq.~\eqref{e4_17b} does not change the linearized form of \eqref{e4_17a}; 
its role is to guarantee that the $s$ components of vector $\vecQ$ equal
their prescribed values {\em exactly} rather than in the linear approximation.

To our knowledge, the multi-component ITEM \eqref{e4_17} has not been presented
in the literature before; however, its close ``relative" (related to a family of
squared-operator methods) was stated in \cite{SOM}. The convergence conditions for 
\eqref{e4_17} are \cite{vecITEM}: (i) the $s\times s$ Jacobian \ 
$\partial \vecQ / \partial \vecmu$ \ must be nonsingular, and \ (ii) the number of
positive eigenvalues of this Jacobian must equal the number of positive eigenvalues
of the linearized operator ${\bf L}$ of Eq.~\eqref{e4_01}. 
These conditions are a counterpart of \eqref{e3_05}.
They guarantee that the linearized operator,
\be
{\mathcal L}{\bf \tu}_n  \equiv {\bf L \tu}_n - 
\Ucal \, \langle {\bf N}^{-1} \Ucal,\, \Ucal \rangle^{-1}\,
  \langle {\bf N}^{-1}\Ucal, \, {\bf L} {\bf \tu}_n \rangle
\label{e4_18}
\ee
of the expression $\Lcalnot {\bf u}_n$ appearing inside the parentheses
in \eqref{e4_17a}, is negative definite on the space of
vector functions satisfying an analog of the orthogonality relation \eqref{e3_03}:
\be
\langle {\Ucal,\, \bf \tu}_n \rangle = \vec{0}\,.
\label{e4_19}
\ee

The corresponding multi-component version of the modified CGM \eqref{e3_14} is:
\bsube
\be
\vecr_0= \vecN^{-1} \Lcalnot {\bf u}_0, \qquad 
\vecd_0=\vecr_0 -  \Ucal_0\, \langle \Ucal_0,\, \Ucal_0 \rangle^{-1}\, 
   \langle  \Ucal_0,\, \vecr_0 \rangle\,,
\label{e4_20a}
\ee
\be
\alpha_n=-
\frac{ \langle \vecr_n,\,\vecN \vecd_n \rangle }
{ \langle \vecd_n,\,\Lcal \vecd_n \rangle }\,,
\label{e4_20b}
\ee
\be
\hat{{\bf u}}_{n+1}={\bf u}_n+\alpha_n \vecd_n\,, \qquad
u^{(k)}_{n+1}=\hat{u}^{(k)}_{n+1} 
\sqrt{ \frac{ Q^{(k)}- \sum_{l=k+1}^S q^{(kl)} \hat{P}^{(l)}_{n+1} }{ \hat{P}^{(k)}_{n+1} } }\;, 
\qquad    k=1,\ldots\,,\, s\le S\,,
\label{e4_20c}
\ee
\be
\vecr_{n+1}= \vecN^{-1} \Lcalnot {\bf u}_{n+1}
\label{e4_20d}
\ee
\be
\beta_n= -
 \frac{ \dst \langle \vecr_{n+1},\,\Lcal \vecd_n \rangle - 
   \langle \Lcal \vecd_n,\, \Ucal_{n+1} \rangle 
   \langle \Ucal_{n+1},\, \Ucal_{n+1} \rangle^{-1}\,
    \langle \Ucal_{n+1}, \, \vecr_{n+1} \rangle  }
{ \langle \vecd_n,\,\Lcal \vecd_n \rangle },
\label{e4_20e}
\ee
\be
\vecd_{n+1}= \vecr_{n+1}+\beta_n \vecd_n - 
 \Ucal_{n+1} \, \langle \Ucal_{n+1},\, \Ucal_{n+1} \rangle^{-1}\, 
 \langle \Ucal_{n+1},\, \vecr_{n+1}+\beta_n \vecd_n \rangle  \,.
\label{e4_20f}
\ee
\label{e4_20}
\esube
This method is guaranteed to converge to a fundamental solitary wave with a prescribed vector
of conserved quantities $\vecQ$ under the conditions stated before Eq.~\eqref{e4_18}.

\setcounter{equation}{0}
\section{Numerical examples}

Here we compare the performance of the modified versions of the CGM presented in Sections
3--5 with the performance of the corresponding Richardson-type methods accelerated by
the mode elimination (ME) technique \cite{ME,SOM}.

The model equations are \eqref{e1_05} and its
two-component extension:
\be
\ba{l}
\dst \nabla^2 u^{(1)} + V_0(\cos^2x + \cos^2y)u^{(1)} + 
     \left( F^{(1)}(u^{(1)})^2 + F^{(12)} (u^{(2)})^2 \right) u^{(1)} = \mu^{(1)} u^{(1)} \vspace{0.2cm} \\
\dst \nabla^2 u^{(2)} + V_0(\cos^2x + \cos^2y)u^{(2)} + 
     \left( F^{(12)}(u^{(1)})^2 + F^{(2)} (u^{(2)})^2 \right) u^{(2)} = \mu^{(2)} u^{(2)}
\ea
\label{e6_01}
\ee
where the constants
$$
F^{(1)}=1, \quad F^{(2)}=4, \quad F^{(12)}=0.5\,.
$$
Note that Eq.~\eqref{e1_05} is equivalent to Eqs. (2.1), (2.2) of \cite{ShiY07}, where
$\mu_{\cite{ShiY07}}=2V_0-\mu_{\rm this\;paper}$. 
The values of $V_0$ and the propagation constant(s) or power(s) are specified below, so that
each modified CGM and the corresponding Richardson-type method accelerated by ME are tested
for three cases: mildly numerically stiff, stiffer, and stiffest. (In the stiffest case the
methods take the most number of iterations to converge.) 
A code for the most complicated of these methods --- the modified CGM \eqref{e2_14}, \eqref{e4_08} for
the two-component Eqs.~\eqref{e6_01} --- is presented in Appendix 3. (A counterpart of this code
for a single-component equation is considerably simpler and can be easily written based on a
code from Appendix 1 in \cite{gP}. Codes seeking solitary waves with prescribed power are also
much simpler than the code in Appendix 3 below, because they do not compute quantities ${\bf N}$,
${\bf e}^{(k)}$, etc.)

We begin by describing the simulations setup for the methods where 
the propagation constant(s) is (are) specified.
We compare the modified CGMs \eqref{e2_14} 
(for the single equation \eqref{e1_05}) and \eqref{e2_14}, 
\eqref{e4_08} (for the two-component system \eqref{e6_01})
with the ME-accelerated generalized \PM. The latter method for a single equation is \cite{ME}:
\bsube
\be
u_{n+1}=u_n + \left[ N^{-1} \Lnot u_n - 
 \gamma  \frac{\langle u_n,\, \Lnot u_n \rangle}{\langle u_n,\, N u_n \rangle}\,u_n
      \, - \, \gamma^{\rm slow}_n 
      \frac{ \langle \phi^{\rm slow}_n,\, \Lnot u_n \rangle }
            { \langle \phi^{\rm slow}_n,\, N \phi^{\rm slow}_n \rangle}\,
      \phi^{\rm slow}_n \,  \right] \D\tau\,;
\label{e6_02a}
\ee
which for a multi-component system extends to:
\be
{\bf u}_{n+1}={\bf u}_n + \left[ {\bf N}^{-1} {\bf \Lnot}{\bf u}_n - 
 \sum_{k=1}^S \gamma^{(k)} 
 \frac{ \left\langle {\bf e}^{(k)}_n,\, {\bf \Lnot}{\bf u}_n \right\rangle }
      { \left\langle {\bf e}^{(k)}_n,\,{\bf N} {\bf e}^{(k)}_n \right\rangle}\,
      {\bf e}^{(k)}_n \, - \, \gamma^{\rm slow}_n
      \frac{ \left\langle {\bf \Phi}^{\rm slow}_n,\, {\bf \Lnot}{\bf u}_n \right\rangle }
            { \left\langle {\bf \Phi}^{\rm slow}_n,\,{\bf N} {\bf \Phi}^{\rm slow}_n \right\rangle}\,
      {\bf \Phi}^{\rm slow}_n \,  \right] \D\tau\,.
\label{e6_02b}
\ee
\label{e6_02}
\esube
Here the notations are as in \eqref{e4_02} above and in addition,
\be
{\bf \Phi}^{\rm slow}_n = {\bf u}_{n} - {\bf u}_{n-1}, \qquad
\gamma^{\rm slow}_n=1+\frac{h}{\lambda^{\rm slow}_n \D\tau}, \qquad
\lambda^{\rm slow}_n = 
\frac{ \left\langle {\bf \Phi}^{\rm slow}_n,\,{\bf L} {\bf \Phi}^{\rm slow}_n \right\rangle}
     { \left\langle {\bf \Phi}^{\rm slow}_n,\,{\bf N} {\bf \Phi}^{\rm slow}_n \right\rangle}\,,
\label{e6_03}
\ee
where $h$ is the fraction of the slowest-decaying mode that is subtracted at each iteration.
In \cite{ME} we advocated subtracting around 70\% of such a mode for robust and
nearly optimal performance; accordingly, here we use $h=0.7$ in all examples.

\begin{figure}[h]
\vspace*{-4cm}
\rotatebox{0}{\resizebox{10cm}{13cm}{\includegraphics[0in,0.5in]
 [8in,10.5in]{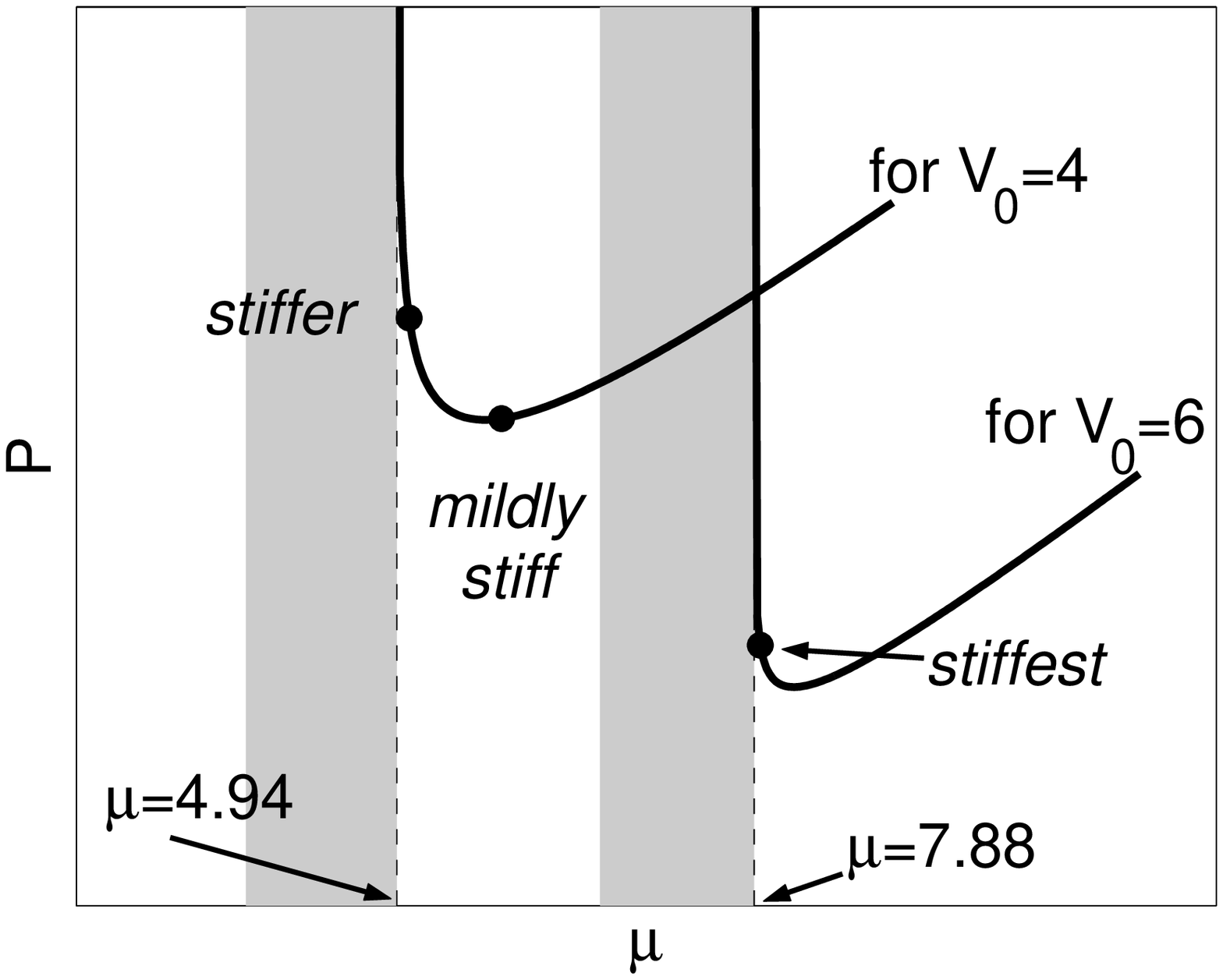}}}
\vspace{-2cm}
\caption{Schematics of the $P(\mu)$ curves for fundamental solitary waves 
of Eq.~\eqref{e1_05} with $V_0=4$ and $V_0=6$. 
The axes, markers, etc. are drawn not to scale.
The three cases of increased
numerical stiffness, listed in \eqref{e6_04a}, 
are labeled with filled circles. The shaded areas represent the first
spectral bands of the linearized operator $L$ for the two values of $V_0$.
}
\label{fig_1}
\end{figure}

\begin{figure}[h]
\vspace*{-4cm}
\rotatebox{0}{\resizebox{10cm}{13cm}{\includegraphics[0in,0.5in]
 [8in,10.5in]{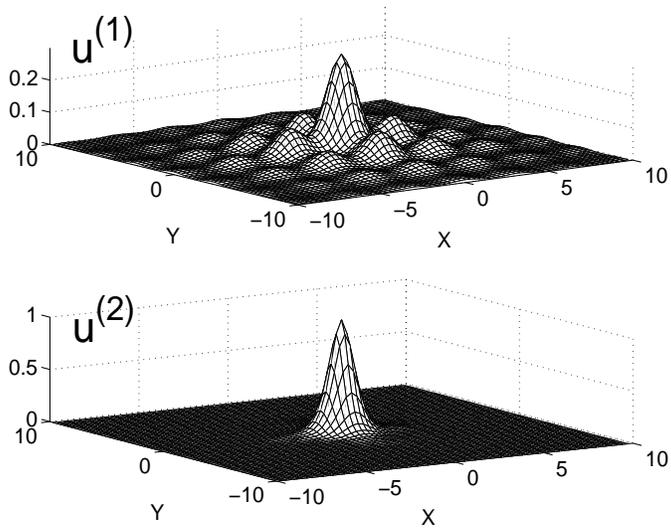}}}
\vspace{-2cm}
\caption{The two-component solution of the stiffer case in \eqref{e6_04b} for Eq.~\eqref{e6_01}.
Note the different vertical scales of the two components.
}
\label{fig_2}
\end{figure}

For all the methods listed in the previous paragraph, we designate the following sets 
of parameters in Eqs.~\eqref{e1_05} and \eqref{e6_01} as corresponding to the mildly stiff,
stiffer, and stiffest cases.
\bsube
\be
\ba{lrcl} 
\mbox{\underline{For \eqref{e1_05}}}: \qquad & 
\mbox{mildly stiff} & \so & V_0=4, \quad \mu=5.03; \vspace{0.1cm} \\
 & \mbox{stiffer} & \so & V_0=4, \quad \mu=4.95; \vspace{0.1cm} \\
 & \mbox{stiffest} & \so & V_0=6, \quad \mu=7.89; 
\ea
\label{e6_04a}
\ee
\be
\ba{lrcl} 
\mbox{\underline{For \eqref{e6_01}}}: \qquad & 
\mbox{mildly stiff} & \so & V_0=4, \quad \mu^{(1)}=5.03, \;\; \mu^{(2)}=5.5; \vspace{0.1cm} \\
 & \mbox{stiffer} & \so & V_0=4, \quad \mu^{(1)}=4.95,\;\;\mu^{(2)}=6.5; \vspace{0.1cm} \\
 & \mbox{stiffest} & \so & V_0=6, \quad \mu^{(1)}=7.89, \;\; \mu^{(2)}=8.5\,.
\ea
\label{e6_04b}
\ee
\label{e6_04}
\esube
Figure \ref{fig_1} shows schematically the power-versus-$\mu$ plots for Eq.~\eqref{e1_05},
with the three cases of \eqref{e6_04a} labeled. Figure \ref{fig_2} shows the solution for
the stiffer case for Eqs.~\eqref{e6_01}; note the vastly different amplitudes and widths 
of the two components. In the other two cases the solution looks qualitatively the same.
The obtained solutions of the single Eq.~\eqref{e1_05} look qualitatively as the first
component of the solution in Fig.~\ref{fig_2}. 
In general, the closer the propagation constant to the edge of
the band gap, the broader and lower the corresponding solution.

For both the modified CGMs and the ME-accelerated generalized \PM s, we start the iterations by the
non-accelerated generalized \PM, and when the error, defined as
\be
\varepsilon_n= \sum_{k=1}^S 
\frac{ \left\langle ({\bf \Lnot u}_n)^{(k)}, \; ({\bf \Lnot u}_n)^{(k)} \right\rangle }
     { \left\langle u_n^{(k)}, \, u_n^{(k)} \right\rangle }\,,
\label{e6_05}
\ee
reaches a threshold 
\be
\varepsilon_{\rm acceleration\; threshold} = 5\cdot 10^{-2},
\label{e6_06}
\ee
we begin the acceleration by either the modified CGM or ME and carry on the iterations until
the error reaches $10^{-10}$. The parameters of ${\bf N}$, ${\bf e}^{(k)}$, $\lambda^{(k)}$,
and $\gamma^{(k)}$ (or their single-component counterparts) 
stop being computed at the threshold \eqref{e6_06} and the latest computed
values of these parameters are used from that moment on. (As was demonstrated in \cite{gP}, 
for Eqs.~\eqref{e6_01} the eigenvector ${\bf e}^{(2)}$ of ${\bf N}^{-1}{\bf L}$ is computed
not too accurately. Yet, we show below that this does not prevent the modified CGM 
\eqref{e2_14}, \eqref{e4_08}
from performing quite well.)

In regards to the specific numeric value on the right-hand side of \eqref{e6_06}, we observed
that the accelerated methods converge in all cases when this value is not too high (say, is less
than $10^{-1}$). 
If the error is substantially higher, the computed parameters of ${\bf N}$ etc. may be too
inaccurate, so that the corresponding operator ${\bf K^{(\rm 0,mod)}}$ would be ``too far" from
self-adjoint, which might then prevent the convergence of algorithms \eqref{e2_14} or \eqref{e4_08}.
In addition, {\em only} for the modified CGM {\em in the stiffest case}, the value 
of the acceleration threshold should not be too low (e.g., 
should be higher than about $5\cdot 10^{-3}$); otherwise, the modified CGM in this case
would diverge. The explanation of this fact was sketched in the last paragraph of Section 3.1.
Here we present specific details for it; 
for the sake of convenience, we do it 
for the single Eq.~\eqref{e1_05} and use the transformed variables \eqref{e2_08}, \eqref{e2_13}. 

First, as is well known (see, e.g., \cite{ShiY07} and references therein), 
when $\mu$ in \eqref{e1_05} is very close to the edge of the spectral gap of the linear part of
$K^{(0)}$ (i.e. in the ``stiffest" case considered here), the shape of the 
exact solution of \eqref{e1_05} is very similar to that of the eigenmode(s) of $K$
with negative near-zero eigenvalue(s).
In particular, the solution is broad, as illustrated in the top panel of Fig.~\ref{fig_2}. 
Next, from (\ref{e2_14}a,b) and an analog of 
\eqref{e1_09} it follows that
at the first CGM iteration, the step size $\alpha_0$ along the search direction $d_0$ is
\be
\alpha_0 = \frac{ \langle K^{(\rm 0, mod)}v_0, \, K^{(\rm 0, mod)}v_0 \rangle }
                { \langle K^{(\rm mod)} \,K^{(\rm 0, mod)}v_0, \, K^{(\rm 0, mod)}v_0 \rangle } 
         \approx  \frac{ \langle K^{(\rm mod)} \, \tv_0, \, K^{(\rm mod)}\,\tv_0 \rangle }
                { \langle (K^{(\rm mod)})^2 \,\tv_0, \, K^{(\rm mod)}\, \tv_0 \rangle }\, \gg 1\,.
\label{e6_07}
\ee
(Here the iteration's number is referenced from the start of the CGM algorithm.)
The last strong inequality follows from two facts: (i) operator $K\;(=N^{-1/2}LN^{-1/2})$ 
has a band of negative eigenvalues
that comes very close to zero (see, e.g., Fig.~4(a) in \cite{YangC06} for a related problem),
and \ (ii) it is precisely a superposition of the corresponding eigenmodes
that dominates the error $\tv_0$ obtained by the generalized \PM\ after sufficiently many iterations
(because these modes decay the slowest in Richardson-type methods; see, e.g., \cite{SOM}). 
Having $\alpha_0\gg 1$ and hence
a large first step in the CGM adds to $v_0$ some superposition of near-zero eigenmodes of $K$
and thereby makes the solution at the next iteration, $v_1$, more broad and hence
even more resembling the near-zero eigenmodes. Then the
step size $\alpha_1$, computed at the next iteration, becomes even greater than $\alpha_0$, and the
solution at the following iteration, $v_2$, is made even more broad. This quickly leads to divergence of the
iterations. (Practically, the solution converges to a nonlocalized two-dimensional plane wave.)
On the other hand, if the acceleration by the modified CGM starts when the error $\tv_0$ is not
too small and hence still contains a significant contribution from the eigenmodes whose eigenvalues
are not too close to zero, the corresponding step size $\alpha_0$ is not too large, and subsequent
CGM iterations are able to gradually reduce the error $\tv_n$. 

The above consideration explains why the CGM iterations
should start when the iteration error is not too low. In practice, however, this is not a limitation,
since in numerically stiff cases, one wants to start the acceleration by
the CGM sooner rather than later since this
considerably reduces the computational time, as we will demonstrate below. 

The initial condition in all cases that we consider is taken as
\bsube
\be
u_0 = 1.5\cdot e^{-(x^2+y^2)}\cdot(1+\epsilon_x x+\epsilon_y y), 
\qquad \epsilon_x=0.1, \quad \epsilon_y=-0.2
\label{e6_08a}
\ee
for Eq.~\eqref{e1_05} and as
\be
\left( \ba{c} u^{(1)} \vspace{0.1cm} \\ u^{(2)} \ea \right)_0 \,=\, 
\left( \ba{c} 0.8 \vspace{0.1cm} \\ 1.5 \ea \right)\; \cdot e^{-(x^2+y^2)}\cdot(1+\epsilon_x x+\epsilon_y y), 
\qquad \epsilon_x=0.1, \quad \epsilon_y=-0.2
\label{e6_08b}
\ee
\label{e6_08}
\esube
for Eqs.~\eqref{e6_01}. The exact values of the amplitude(s) and width of the initial condition
are not essential; all methods converge for a wide range of these parameters. As for the asymmetry
parameters $\epsilon_x$ and $\epsilon_y$, these can be also varied quite a bit as long as the 
shape if the initial condition resembles a pulse with one main peak. However, these parameters
should {\em not} be set to zero (or too close to zero) for the modified CGMs \eqref{e2_14} and \eqref{e4_08}
{\em in the stiffest case}. 
(That is, in the other two, less stiff, cases, 
the modified CGMs converge even for a symmetric initial condition.)
The reason for this is similar to the reason why the acceleration
threshold in \eqref{e6_06}  should not be chosen too small. Namely, iterations of the generalized \PM\
which start with a symmetric initial condition, i.e., \eqref{e6_07} with $\epsilon_x=\epsilon_y=0$, 
produce symmetric solutions at all subsequent iterations. These solutions increasingly resemble
the near-zero eigenmode of the linearized operator $L$. Then when the CGM starts, the first
step $\alpha_0$ along the search direction $d_0$ becomes too large, which leads to divergence of
the iterations, as explained after Eq.~\eqref{e6_07}. On the other hand, having an asymmetric initial
condition leads to the iteration error being asymmetric, and such an asymmetric error can have
a considerable content of eigenmodes of $L$ whose eigenvalues are not close to zero. 
This reduces the step size $\alpha_0$ at the first CGM iteration, and
the modified CGM is able to converge.
Note that from the practical perspective, starting with an asymmetric initial condition is not
a limitation of the method. 

While the ME-based acceleration does not require an asymmetric initial condition for convergence,
we still use the same
expressions \eqref{e6_08} for all methods, so as to make our performance comparison uniform. 

We now comment on the choices of the ``fictitious time" step size $\D\tau$. For each simulation,
we first selected its nearly optimal value for the corresponding non-accelerated generalized \PM.
This takes just a couple of trials since the optimal $\D\tau$ is only slightly less than the maximum
value of this parameter for which the generalized \PM\ still converges. (See, e.g., Eqs. (2.5) and (2.6)
in \cite{ME} and Fig.~1(d) in \cite{SOM}, based on the same equations, although for another
iterative method.) Once this $\D\tau_{\rm opt}$ has been determined, we use a slightly smaller
(specifically, by 0.1) value of the step size for the method accelerated by ME; 
the analysis of ME \cite{ME} predicts that this should lead to a more robust and smooth peformance
than using $\D\tau_{\rm opt}$. For example, if we find that $\D\tau_{\rm opt}=0.9$ for the 
non-accelerated generalized \PM, we then use $\D\tau=0.8$ for the ME-accelerated version of this
method. For the method accelerated by a modified CGM, we also start
the iterations using $\D\tau=\D\tau_{\rm opt}-0.1$,
although this does not noticeably affect its performance
(since the CGM itself does not involve $\D\tau$).

Finally, the domain for all our numerical simulations is $[-6\pi,\,6\pi]\times [-6\pi,\,6\pi]$,
with $2^8 \times 2^8$ grid points. 

\begin{figure}[h]
\rotatebox{0}{\resizebox{10cm}{13cm}{\includegraphics[0in,0.5in]
 [8in,10.5in]{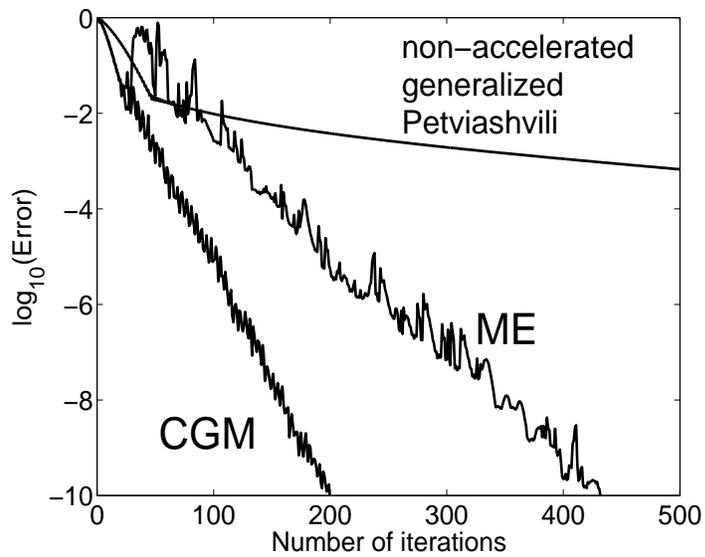}}}
\vspace{-2cm}
\caption{Evolutions of the iteration error, defined by \eqref{e6_05} with $S=1$, for the
stiffest case \eqref{e6_04a} for Eq.~\eqref{e1_05}. The curve for the non-accelerated generalized
\PM\ \eqref{e2_04} is not shown in full so as not to obscure the details of the other two curves.
}
\label{fig_3}
\end{figure}

Tables 1 and 2 list the numbers of iterations and time (both rounded to the nearest ten) 
for each of the three methods
(non-accelerated generalized \PM\ and its two versions accelerated by ME and CGM) in the three cases
\eqref{e6_04} for Eqs.~\eqref{e1_05} and \eqref{e6_01}. In accordance to the note above, the
value of $\D\tau$ is listed only for the non-accelerated method. In Fig.~\ref{fig_3} we plot the
iteration error \eqref{e6_05} versus the iteration number for the stiffest case for Eq.~\eqref{e1_05}.
The error evolutions shown there are representative of all the other cases, except that in the less
stiff cases, the curves corresponding to the ME and CGM are less jagged.

\begin{table}
\renewcommand{\arraystretch}{1.25}
\begin{tabular}{c|c|c|c|} \cline{2-4}
 & \multicolumn{3}{c|}{Iterative method} \\ \hline 
 \multicolumn{1}{|c||}{Case in \eqref{e6_04a}} & 
 non-accelerated \eqref{e2_04} & accel. by ME \eqref{e6_02a} &  accel. by CGM \eqref{e2_14} \\ \hline\hline
 \multicolumn{1}{|c||}{mildly stiff, $\D\tau=1.1$} &  300 iterations, 130 s &
                                                    110 iterations, 60 s & 60 iteratons, 40 s \\ \hline
 \multicolumn{1}{|c||}{stiffer, $\D\tau=1.1$} &  920 iterations, 390 s &
                                                    290 iterations, 170 s & 100 iteratons, 60 s \\ \hline
 \multicolumn{1}{|c||}{stiffest, $\D\tau=1.0$} &  3700 iterations, 1560 s &
                                                    430 iterations, 240 s & 200 iteratons, 120 s \\ \hline                                                    
\end{tabular}
\caption{Comparative performance of the generalized \PM\ accelerated by ME \eqref{e6_02a}
         and the modified CGM \eqref{e2_14} for the single-component Eq.~\eqref{e1_05}.}
\end{table}
\begin{table}
\renewcommand{\arraystretch}{1.25}
\begin{tabular}{c|c|c|c|} \cline{2-4}
 & \multicolumn{3}{c|}{Iterative method} \\ \hline 
 \multicolumn{1}{|c||}{Case in \eqref{e6_04b}} & 
 non-accelerated \eqref{e4_02} & accel. by ME \eqref{e6_02b} &  accel. by CGM \eqref{e4_08} \\ \hline\hline
 \multicolumn{1}{|c||}{mildly stiff, $\D\tau=1.0$} &  330 iterations, 310 s &
                                                    120 iterations, 130 s & 70 iteratons, 70 s \\ \hline
 \multicolumn{1}{|c||}{stiffer, $\D\tau=1.0$} &  780 iterations, 740 s &
                                                    200 iterations, 220 s & 130 iteratons, 130 s \\ \hline
 \multicolumn{1}{|c||}{stiffest, $\D\tau=0.9$} &  3330 iterations, 3110 s &
                                                    550 iterations, 610 s & 240 iteratons, 250 s \\ \hline                                                    
\end{tabular}
\caption{Comparative performance of the generalized \PM\ accelerated by ME \eqref{e6_02b}
         and the modified CGM \eqref{e2_14}, \eqref{e4_08} for the two-component Eq.~\eqref{e6_01}.}
\end{table}

We also verified that a straightforward extension of the CGM, as described at the end of Section 3.1, 
would diverge for all cases listed above and even for non-stiff cases (not listed here).

From these tables we see that, as expected, both ME- and CGM-based accelerations considerably improve the
performance of the iterative method, with the stiffer the case, the more the improvement. 
Furthermore, the CGM-based acceleration performs better than the ME-based one by a factor of 2--2.5;
again, the stiffer the case is, the more improvement the CGM provides over the ME.

\bigskip

We now present similar results when the power \eqref{e1_27}, or its multi-component counterpart
\eqref{e4_09}, of the solution is specified. The non-accelerated methods are the
ITEMs \eqref{e3_02} and \eqref{e4_17} applied to the single- and multi-component Eqs.~\eqref{e1_05}
and \eqref{e6_01}, respectively. 
When the ME-based acceleration is applied, the lines updating
the intermediate solutions $\hat{u}_{n+1}$ and $\hat{\bf u}_{n+1}$ of these methods become:
\bsube
\be
\hat{u}_{n+1}=u_n+ \left[ N^{-1} \left( \Lnnot u_n - \mu_n u_n \right) \,-\,
 \gamma^{\rm slow}_n
       \frac{ \langle \phi^{\rm slow}_n,\, \Lnot u_n \rangle }
             { \langle \phi^{\rm slow}_n,\, N \phi^{\rm slow}_n \rangle}\,
       \phi^{\rm slow}_n \,  \right] \D\tau\,;
\label{e6_09a}
\ee
\begin{multline}
\hat{\bf u}_{n+1}={\bf u}_n + \left[ {\bf N}^{-1} \left( {\bf \Lnnot u}_n - 
 \Ucal_n \, \langle {\bf N}^{-1} \Ucal_n,\, \Ucal_n \rangle^{-1}\,
  \langle {\bf N}^{-1} \Ucal_n, \, {\bf \Lnnot} {\bf u}_n \rangle  \right) \right. \\
  \left. {}- \;  \gamma^{\rm slow}_n 
        \frac{ \left\langle {\bf \Phi}^{\rm slow}_n,\, {\bf \Lnot}{\bf u}_n \right\rangle }
              { \left\langle {\bf \Phi}^{\rm slow}_n,\,{\bf N} {\bf \Phi}^{\rm slow}_n \right\rangle}\,
      {\bf \Phi}^{\rm slow}_n \,  \right] \, \D\tau\,,
\label{e6_09b}
\end{multline}
\label{e6_09}
\esube
where $\gamma^{\rm slow}_n$ and ${\bf \Phi}^{\rm slow}_n$ are defined as in \eqref{e6_03}
and ${\bf \Lnot u}_n$ is the term in parentheses in \eqref{e6_09b};
the last lines remain the same as in the respective methods \eqref{e3_02} and \eqref{e4_17}.
The modified CGMs for the single- and two-component equations are \eqref{e3_14} and \eqref{e4_20}.

The three cases of increasing numerical stiffness are chosen as,
\bsube
\be
\ba{lrcl} 
\mbox{\underline{For \eqref{e1_05}}}: \qquad & 
\mbox{mildly stiff} & \so & V_0=4, \quad P=2.1\; (\mu=5.08); \vspace{0.1cm} \\
 & \mbox{stiffer} & \so & V_0=4, \quad P=1.94 \; (\mu=5.01); \vspace{0.1cm} \\
 & \mbox{stiffest} & \so & V_0=6, \quad P=0.92\;(\mu=7.93); 
\ea
\label{e6_10a}
\ee
\be
\ba{lrcl} 
\mbox{\underline{For \eqref{e6_01}}}: \qquad & 
\mbox{mildly stiff} & \so & V_0=4, \quad P^{(1)}=1.50\;(\mu^{(1)}=5.10), \;\; 
                                         P^{(2)}=1.00\; (\mu^{(2)}=5.92); \vspace{0.1cm} \\
 & \mbox{stiffer} & \so & V_0=4, \quad P^{(1)}=0.50\;(\mu^{(1)}=4.98), \;\; 
                                       P^{(2)}=1.50\; (\mu^{(2)}=6.62); \vspace{0.1cm} \\
 & \mbox{stiffest} & \so & V_0=6, \quad P^{(1)}=0.49\;(\mu^{(1)}=7.93), \;\; 
                                        P^{(2)}=0.60\; (\mu^{(2)}=8.55)\,,
\ea
\label{e6_10b}
\ee
\label{e6_10}
\esube
where the values of $\mu$, as computed within the methods, are listed for reference only. 
Note that for the single-component equation, we had to choose the parameters for which the solution
would satisfy the stability condition \eqref{e3_05} (see Fig.~\ref{fig_1})
of the iterative methods \eqref{e3_02},
\eqref{e6_09a}, and \eqref{e3_14}. For the two-component equation, the parameters were chosen
by trial and error. 

The preconditioning operator was taken to be of the form \eqref{e1_24} or, in the two-component case, as
\be
{\bf N} = \left( c-\nabla^2 \right) \,{\rm diag}(1,\,1)
\label{e6_11}
\ee
with $c=1$ (which did not necessarily result in the optimal convergence rate). 
The ``fictitious time" step $\D\tau$ for the ME-based methods was chosen as
$\D\tau=\D\tau_{\rm opt}-0.1$, as described above, where $\D\tau_{\rm opt}$ was the
optimal step size for the non-accelerated ITEM.
The computational domain 
and the initial conditions were taken as for the
methods that find solutions with prescribed values of the propagation constant.
Note that here, the modified CGMs \eqref{e3_14} and \eqref{e4_20} are guaranteed to
converge to the solution for {\em any} initial conditions that resemble a two-dimensional pulse.
Nonetheless we used \eqref{e6_08} with $\epsilon_x=0.1$ and $\epsilon_y=-0.2$ just for uniformity
of all our simulations. For the same reason, we also started the acceleration 
at the same threshold \eqref{e6_06}, even though both the ME-based methods \eqref{e6_09} and the 
modified CGMs \eqref{e3_14} and \eqref{e4_20} could be employed at the very first iteration. 
The results are summarized in Tables 3 and 4.

\begin{table}
\renewcommand{\arraystretch}{1.25}
\begin{tabular}{c|c|c|c|} \cline{2-4}
 & \multicolumn{3}{c|}{Iterative method} \\ \hline 
 \multicolumn{1}{|c||}{Case in \eqref{e6_10a}} & 
 non-accelerated \eqref{e3_02} & accel. by ME \eqref{e6_09a} &  accel. by CGM \eqref{e3_14} \\ \hline\hline
 \multicolumn{1}{|c||}{mildly stiff, $\D\tau=0.9$} &  330 iterations, 140 s &
                                                    90 iterations, 50 s & 50 iteratons, 40 s \\ \hline
 \multicolumn{1}{|c||}{stiffer, $\D\tau=1.0$} &  1670 iterations, 710 s &
                                                    160 iterations, 100 s & 120 iteratons, 90 s \\ \hline
 \multicolumn{1}{|c||}{stiffest, $\D\tau=0.6$} &  4690 iterations, 1980 s &
                                                    550 iterations, 330 s & 210 iteratons, 150 s \\ \hline                                                    
\end{tabular}
\caption{Comparative performance of the ITEM accelerated by ME \eqref{e6_09a}
         and the modified CGM \eqref{e3_14} for the single-component Eq.~\eqref{e1_05}.}
\end{table}
\begin{table}
\renewcommand{\arraystretch}{1.25}
\begin{tabular}{c|c|c|c|} \cline{2-4}
 & \multicolumn{3}{c|}{Iterative method} \\ \hline 
 \multicolumn{1}{|c||}{Case in \eqref{e6_10b}} & 
 non-accelerated \eqref{e4_17} & accel. by ME \eqref{e6_09b} &  accel. by CGM \eqref{e4_20} \\ \hline\hline
 \multicolumn{1}{|c||}{mildly stiff, $\D\tau=0.6$} &  300 iterations, 260 s &
                                                    120 iterations, 150 s & 60 iteratons, 80 s \\ \hline
 \multicolumn{1}{|c||}{stiffer, $\D\tau=0.6$} &  850 iterations, 740 s &
                                                    220 iterations, 280 s & 120 iteratons, 130 s \\ \hline
 \multicolumn{1}{|c||}{stiffest, $\D\tau=0.5$} &  1610 iterations, 1400 s &
                                                    380 iterations, 480 s & 130 iteratons, 160 s \\ \hline                                                    
\end{tabular}
\caption{Comparative performance of the ITEM accelerated by ME \eqref{e6_09b}
         and the modified CGM \eqref{e4_20} for the two-component Eq.~\eqref{e6_01}.}
\end{table}

Again, as expected, we see that both ME- and CGM-based accelerations considerably improve the
performance of the iterative method, with the stiffer the case, the more the improvement. 
Furthermore, the CGM-based acceleration performs better than the ME-based one by a factor of 2--3
for the two-component equation, with the stiffer the case, the more imrovement being provided 
by the CGM over the ME. Interestingly, however, for the single-component equation, the modified CGM
provides an improvement over the ME only for the stiffest case; for somewhat less stiff cases, 
the improvement (in terms of computing time) is only marginal.


\section{Conclusions}

In this work, we proposed modifications of the well-known Conjugate Gradient method (CGM)
that are capable of finding fundamental solitary waves of single- and multi-component
Hamiltonian nonlinear equations.
Our modified CGMs converge much faster than previously considered ierative methods of 
Richardson's type, like the (generalized) Petviashvili and imaginary-time evolution methods.
Moreover, the slower the convergence of the Richardson-type method, the more speedup the modified
CGM provides.

The classic CGM for a linear system of equations is known
to converge when the matrix of this system
is sign definite. For most solitary waves, the linearized operator about them,
which is a counterpart of the matrix in the previous sentence, is not sign definite.
While, in theory, this does not automatically imply that a straightforward generalization
of the CGM to stationary nonlinear wave equations should fail (e.g., diverge), we verified
that for the equations considered in Section 6, it does indeed fail. Then, the thrust of this
work was to modify the CGM in such a way that it would be guaranteed to converge even when
the linearized operator has eigenvalues of either sign. More precisely, our goal was to
develop such modified versions of the CGM in the subcase when the number of the positive eigenvalues
of that operator does not exceed the number of the components of the solitary wave.
According to our ``definition" at the end of Introduction, this situation would hold for
fundamental solitary waves.

The modified CGMs that we proposed in Sections 3 to 5 do not, strictly speaking, meet that goal.
However, we show below that they come as closely as theoretically possible to doing so.
As far as their practical applications, we
demonstrated that our modified CGMs converge in the same parameter regions as earlier proposed,
slower iterative methods. This convergence can be achieved by a simple shift of the initial
condition, {\em which has a theoretical explanation} given in Section 6. 

When finding solitary waves with prescribed values of the propagation constants, we consider
an equivalent nonlinear equation whose linearized operator is modified in such a way that
its only positive eigenvalue is essentially turned into a negative one; see Eqs. \eqref{e2_15}
and \eqref{e4_08}. This, however, makes the modified linearized operator ``slightly" non-self-adjoint,
because the generalized eigenfunction(s) of the original linearized operator, as in \eqref{e2_09} and
\eqref{e4_04}, is (are) available only approximately. This is a fundamental fact about all nonlinear
equations except for their very narrow subclass, equations with power-law nonlinearity \eqref{e1_01}
(see also their two-component generalization in \cite{gP}), and hence it, in general, cannot be improved.
Thus, this fact causes ``slight" non-self-adjointness of the modified linearized operator. This, in
turn, leads to sign indefiniteness of the quadratic form in (\ref{e2_14}b,e) and thereby prevents
one from obtaining a rigoroius guarantee that the modified CGM would always converge. However, we 
explained in Section 6 and Appendix 1 that the method can diverge {\em only} when $L$ has small 
eigenvalues, and for those cases pointed out that a simple shift of the initial condition will 
suffice to make the modified CGM to actually converge.

For finding solitary waves with prescribed values of the power \eqref{e1_27} or,
more generally, quadratic conserved quantities \eqref{e4_09},
the situation is different. There, the modified CGM is caused to converge not by modifying a linearized
operator but by making the search directions satisfy a certain orthogonality relation; see 
\eqref{e3_03} and \eqref{e4_19}. The linearized operator employed by the method 
can be shown \cite{ITEM}
to be self-adjoint in the space of functions satisfying those orthogonality relations, but it is
negative definite only under conditions stated before Eqs.~\eqref{e3_05} and \eqref{e4_18}.
This restriction is intrinsic to a method seeking a solution with a prescribed power, and
hence cannot be relaxed for any nonlinear wave equations. 

Thus, we have justified why the modified CGMs proposed in this work come as closely as theoretically
possible to guaranteeing that they would converge to fundamental solitary waves. In practice, however,
these methods can always be forced to converge, as have been mentioned above
and demonstrated in Section 6.

The modified CGMs are faster not only than Richardson-type methods, but also than those
methods accelerated by the slowest-decaying mode elimination (ME) technique \cite{ME,SOM}.
Namely, in comparison to the ME-accelerated methods, 
the modified CGMs provide an improvement of about a factor of two to three 
in terms of computing time;
see Tables 1--4 in Section 6 for details.
Importantly, the CGMs do so in the most numerically stiff cases, when the respective non-accelerated
methods would converge extremely slowly and hence the acceleration would be most desirable.
In such cases, it would pay off to use the modified CGMs instead of ME, even though the latter
is a little easier to program into a code. On the other hand, in non-stiff cases, 
i.e. when the non-accelerated
methods would converge in just a few tens of iterations, both the modified CGMs and ME would provide
only modest improvement in computing time; compare \eqref{e1_21} and \eqref{e1_add01}. In such
cases, it may be simpler to use ME or even the non-accelerated method. Let us point out
one other, practical, advantage of the ME-based acceleration over the CGMs. 
Namely, the fact that we were able
to construct modified CGMs is critically grounded in the existence of relations \eqref{e2_15}, 
\eqref{e4_08} or \eqref{e3_03}, \eqref{e4_19}, as we explained above. 
On the contrary, ME can be used to accelerate {\em any}
Richardson-type iterative method, e.g., methods from 
\cite{MusslimaniY04}--\cite{StepanyantsT06} or the ITEM with amplitude normalization \cite{ITEM}.
Of course, while these methods do converge in many cases, their convergence conditions are
not known, and hence their use with or without ME-based acceleration would come without a
guarantee that they would converge.

Finally, let us briefly mention alternative iterative methods, other than Newton's method proper, 
which can be used to compute solitary waves, both fundamental and non-fundamental. 
In a recent numerical study \cite{J09}, Yang showed that a certain combination of the CGM and Newton's 
method converged to both fundamental and non-fundamental solitary waves for all of the examples
considered in that study. 
It is remarkable, and at the moment has no rigorous theoretical explanation, that
the straightforward generalization of the CGM outlined at the end of Section 3.1
would diverge for most of the same 
examples for which Yang's method converges. Thus, Yang's method can be used
in practice; however, it should be kept in mind that it can be possible to
encounter situations where it would diverge.
Alternatively, if one seeks a non-fundamental solitary wave, 
one can ``square" the equation (see (\ref{e1_14}) and (\ref{e1_15}))
and apply the original CGM to it. For linear systems, this trick has long been known (see,
e.g., \cite{TrefethenBau_NLA}, p. 304), and some researchers used it for finding 
non-fundamental solitary waves \cite{Sukhorukov_private}. 
Let us note, however, that ``squaring" an equation 
leads to squaring the condition number of the corresponding linearized operator
(since ${\rm cond}(A^2)=({\rm cond}(A))^2$), which, according to \eqref{e1_add01},
slows down the convergence; also, more arithmetic operatons per iteration are required
for a ``squared" method.

\section*{Acknowledgement}
This work at its early stage was supported in part by the National Science Foundation under
grant DMS-0507429. The author thanks Dr. J. Yang for useful discussions and Mr. N. Jones
for assistance with computations at the early stage of this work.


\setcounter{equation}{0}
\renewcommand{\theequation}{A1.\arabic{equation}}
\section*{Appendix 1: \ Technical results for Sections 3 and 4}

Here we will first prove statements (i)--(iii) found in the first paragraph of Section 3.3 and
also explain the choice \eqref{e2_12b} for the constant $\Gamma$. 
Then we will prove that the fact that $L$ may have a zero eigenvalue due to a translational 
eigenmode will not cause the CGM to break down. Finally, we will supply missing steps in
the derivations of \eqref{e3_12} and \eqref{e3_13}.

From \eqref{e2_12a},
\be
K^{(0,\rm mod)}v=0 \qquad \so \qquad \Knot v = \chi v, \quad 
\chi \,\equiv\, \Gamma \,\frac{ \langle v,\Knot v \rangle}{\langle v, v \rangle }\,.
\label{A1_01}
\ee
Substituting the second equation in \eqref{A1_01} into the definition of $\chi$ we find:
\be
\chi = \Gamma \frac{ \langle v,\chi v \rangle}{\langle v, v \rangle } = \chi\Gamma\,.
\label{A1_02}
\ee
Since $\Gamma\neq 1$ (see \eqref{e2_12b}), Eq.~\eqref{A1_02} implies that $\chi=0$,
which along with the second equation in \eqref{A1_01} shows that \
$(\Knot v=0) \; \Leftrightarrow (K^{(0,\rm mod)} v=0)$. This proves statement (i). 

Statement (ii), i.e. that $K^{(\rm mod)}$ is approximately self-adjoint,
follows by considering the following inner product for arbitrary functions $f$ and $g$: 
\begin{eqnarray}
\left\langle f, K^{(\rm mod)} g \right\rangle & = & 
 \langle f, K g \rangle - \Gamma \frac{ \langle v, K g \rangle }{ \langle v, v \rangle }
  \langle f, v \rangle   \nonumber \\
  & \approx &   \langle f, K g \rangle - 
 \Gamma \frac{ \langle v, \lambda^{(1)} g \rangle \, \langle v, f \rangle    }{ \langle v, v \rangle }
    \nonumber \\
 & \approx &   \langle g, K f \rangle - 
  \Gamma \frac{ \langle v, g \rangle \, \langle v, K f \rangle    }{ \langle v, v \rangle }
    \nonumber \\
 & = & \left\langle g, K^{(\rm mod)} f \right\rangle\,.
\label{A1_03}
\end{eqnarray}
The two approximate equalities above are due to the approximate relation \eqref{e2_09},
and we have also used the fact that $K$ is self-adjoint.

Statement (iii), i.e. that all eigenvalues of $K^{(\rm mod)}$ that are not too close to zero
are negative, follows by analogy with the statement found after Eq.~\eqref{e1_19b}.
First, due to \eqref{e2_09}, one eigenfunction, $\psi^{(1)}$,
of $K^{(\rm mod)}$ is close to $v$, and for it,
\be
K^{(\rm mod)} \psi^{(1)} \approx K^{(\rm mod)} v \approx \lambda^{(1)}(1-\Gamma)v
\approx \lambda^{(1)}(1-\Gamma)\psi^{(1)} \,.
\label{A1_04}
\ee
When $\Gamma$ is chosen according to \eqref{e2_12b}, the eigenvalue 
corresponding to $\psi^{(1)}$ is approximately $-1$, which is near the accumulation point
of the spectrum of $K$ \cite{ITEM}. It is convenient to ``place" this eigenvalue 
inside (as opposed to at the edge of) the spectrum of $K^{(\rm mod)}$ 
because then it does not affect the condition number of the operator
and hence the convergence factor of the iterative method; see \eqref{e1_22}, \eqref{e1_21},
\eqref{e1_add01}. Thus, we have shown that the ``main culprit" of sign indefiniteness of
$K$ --- the eigenvalue $\lambda^{(1)}$ --- has been successfully dealt with.
To finish the proof of statement (iii), let us show that $K^{(\rm mod)}$ can, in principle, acquire
small positive eigenvalue(s). Let $\psi^{(k)},\;k\ge 2$ be eigenfunctions of $K$
with negative eigenvalues.
Since $v$ is only an approximate eigenfunction of the self-adjoint
operator $K$, then $\langle v, K\psi^{(k)} \rangle = \langle Kv, \psi^{(k)} \rangle$ 
does not exactly equal to zero, which causes the eigenfunctions, and hence eigenvalues,
of $K^{(\rm mod)}$ to differ slightly from those of $K$. Hence small negative eigenvalues
of $K$ could, in theory, become small positive eigenvalues of $K^{(\rm mod)}$. This, however, 
does {\em not} appear to be the case in practice, since if it had been, then the generalized \PM\ 
\eqref{e2_04} would diverge (see the text before Eq.~\eqref{e2_add01}),
whereas our numerical experiments conducted in \cite{gP}
and in this paper do not show such a divergence.

Even if $K^{(\rm mod)}$ has only negative eigenvalues, the quadratic form 
$\langle \bar{d}_n,\, K^{(\rm mod)} \bar{d}_n \rangle$ may still not be negative definite
because $K^{(\rm mod)}$ is not self-adjoint. Since, however, $K^{(\rm mod)}$ is only
{\em ``slightly"} non-self-adjoint, the sign indefiniteness of the above quadratic form
can occur only when some eigenvalues of $K^{(\rm mod)}$ are close to zero, as illustrated
by the following $2\times 2$ example:
\be
( 3\epsilon \;\; 1) \left( \ba{rr} -1 & 5\epsilon\hspace*{0.1cm} \\ 0 & -4 \epsilon^2 \ea \right) 
 \left( \ba{c} 3\epsilon \\ 1 \ea \right) = 2\epsilon^2 > 0\,.
\label{A1_05}
\ee
Note that, for $\epsilon\ll 1$, 
the vector in \eqref{A1_05} is aligned primarily with
the eigenvector, $(5\epsilon \;\; 1)^T$, corresponding to the smaller eigenvalue of the matrix.
The practical implication of the sign indefiniteness of the quadratic form 
$\langle \bar{d}_n,\, K^{(\rm mod)} \bar{d}_n \rangle$ is that it, and hence the denominators
in (\ref{e2_14}b,e), can become arbitrarily close to zero. This may cause divergence of the
modified CGM \eqref{e2_14}. 

To conclude the consideration of statement (iii), 
note that the eigenvalues of $K=N^{-1/2}LN^{-1/2}$ are the same as those of $N^{-1}L$.
According to the Sylvester law of inertia \cite{HornJohnson_book}, eigenvalues of $N^{-1}L$ and
$L$ have the same signs. Moreover, for reasonable choices of $N$ (i.e., for $c=O(1)$ in \eqref{e1_24}),
the eigenvalues of both operators are of the same order of magnitude. Thus, small
eigenvalues of $K$ (and hence of $K^{(\rm mod)}$, see above) imply small eigenvalues of $L$.

\smallskip

Let us now show that if operator $L$ has an eigenmode $\psi_{\rm trans}$ corresponding to
translational invariance of the solitary wave, this would not cause a breakdown of algorithm \eqref{e2_14}.
As before, we will perform the analysis in transformed variables \eqref{e2_08} and \eqref{e2_13}
and in the linear approximation with respect to $\tv_n$, $\bar{r}_n$, and $\bar{d}_n$.

The accordingly transformed translational eigenmode satisfies
\be
K^{(\rm mod)}  \bptrans =0\,.
\label{A1_add01}
\ee
Let us also write down the counterparts, respectively numbered,
of the lines of algorithm \eqref{e2_01} that we will refer to:
\bsube
\addtocounter{equation}{3}
\be
\bar{r}_{n+1}=K^{(0,\rm mod)}v_{n+1} = K^{(\rm mod)} \tv_{n+1}\,.
\label{A1_add02d}
\ee
\addtocounter{equation}{1}
\be
\bar{d}_{n+1}=\bar{r}_{n+1}+\beta_n \bar{d}_n\,.
\label{A1_add02f}
\ee
\addtocounter{equation}{-5}
\be
\alpha_{n+1}= - 
 \frac{ \langle \bar{r}_{n+1}, \bar{d}_{n+1} \rangle }{ \langle K^{(\rm mod)} \bar{d}_{n+1}, \bar{d}_{n+1} \rangle}\,.
\label{A1_add02b}
\ee
\label{A1_add02}
\esube
We will show by contradiction
that if $\bar{d}_n$ is not proportional to $\bptrans$ then neither will be $\bar{d}_{n+1}$.
Then, clearly, since all the eigenvalues of $K^{(\rm mod)}$ other than $\bptrans$ are negative,
the denominator in \eqref{A1_add02b} will never vanish and hence the CGM will not break down.
First, the solvability condition (Fredholm alternative) for \eqref{A1_add02d} implies that \
\be
\bar{r}_{n+1}=\bpperp, \qquad \mbox{where $\langle \bpperp, \bptrans \rangle = 0$,}
\label{A1_add03}
\ee
and hence $\beta_n\neq 0$ since $K^{(\rm mod)} \bar{d}_n\neq 0$. 
Second, from \eqref{e2_02} and \eqref{A1_add03} it follows that
\be
\langle \bar{d}_{n}, \bpperp \rangle =0\,.
\label{A1_add04}
\ee
Third, let us assume that for some $n$, $\bar{d}_{n+1}=a\bptrans$. Note that by
\eqref{e2_02} and \eqref{A1_add02f}, $a\neq 0$. 
Then from \eqref{A1_add02f} it follows that
\be
\bar{d}_n=\frac{a\bptrans - \bpperp}{\beta_n} \qquad \so \qquad
\langle \bar{d}_n, \bpperp \rangle = \frac{a}{\beta_n} \langle \bpperp, \bpperp \rangle \neq 0.
\label{A1_add05}
\ee
This contradicts \eqref{A1_add04}, and hence $\bar{d}_{n+1}$ can never become proportional to $\bptrans$,
which proves our assertion. Note that if $\tv_{n+1}$ becomes proportional to $\bptrans$, 
then $\bar{r}_{n+1}=0$, in which case the iterations converge to $(v+{\rm const}\cdot\bptrans)$,
which is just a translated solitary wave $v$.

\smallskip

Finally, let us show that \eqref{e3_12} and \eqref{e3_13} hold. 
Since, as we noted in Section 4.2, the second equation in
\eqref{e3_09c} does not change the linearized form of the first equation, we can use $\hat{v}_{n+1}$
instead of $v_{n+1}$ in \eqref{e3_09d}. Then
\be
\bar{r}_{n+1}=\bar{r}_n+\alpha_n \Kcal \bar{d}_n + O(\tv_n^2)\,.
\label{A1_06}
\ee
Along with \eqref{e3_09b}, this implies \eqref{e3_12}. Next, the only step that requires
some explanation in the derivation of \eqref{e3_13} is the omission of \ 
$\langle \bar{d}_n, \, N^{-1} v_{n+1} \rangle$. By \eqref{e3_10}, this inner product is
\be
\langle \bar{d}_n, \, N^{-1} v_{n+1} \rangle = \langle \bar{d}_n, \, N^{-1} v_{n} \rangle +
O(\tv_n^2) \,= \, O(\tv_n^2)\,,
\label{A1_07}
\ee
so that its omission does not change the accuracy implied by the right-hand side of \eqref{e3_13}.

\setcounter{equation}{0}
\renewcommand{\theequation}{A2.\arabic{equation}}
\section*{Appendix 2: \ An alternative modified CGM for Section 3}

Since, for the reasons explained at the end of Section 3,
we do not employ this algorithm for numerical examples presented in this paper,
below we will present it only in terms of the transformed variables defined in
\eqref{e2_08} and \eqref{e2_13}.

\bsube
\be
\bar{r}_0= \Knot v_0, \qquad 
\bar{d}_0=\bar{r}_0 - \frac{\langle \bar{r}_0,v_0 \rangle}{\langle v_0,v_0 \rangle}v_0,
\label{e2_16a}
\ee
\be
\alpha_n=-
\frac{ \dst \langle \bar{r}_n,\,\bar{d}_n \rangle - 
  \frac{ \langle v_n, K \bar{d}_n \rangle \langle \bar{r}_n,\,v_n \rangle}
       { \lambda^{(1)} \langle v_n,\,v_n \rangle } }
{ \langle \bar{d}_n,\,K\bar{d}_n \rangle },
\label{e2_16b}
\ee
\be
v_{n+1}=v_n+\alpha_n \bar{d}_n - 
   \frac{ \langle v_n,\,\Knot v_n \rangle}{ \lambda^{(1)} \langle v_n,\,v_n \rangle }v_n ,
\label{e2_16c}
\ee
\be
\bar{r}_{n+1}= \Knot v_{n+1}
\label{e2_16d}
\ee
\be
\beta_n= -
 \frac{ \dst \langle \bar{r}_{n+1},\,K\bar{d}_n \rangle - 
   \frac{ \langle v_n, K \bar{d}_n \rangle \langle \bar{r}_{n+1},\,v_{n+1} \rangle}
        { \langle v_{n+1},\,v_{n+1} \rangle } }
{ \langle \bar{d}_n,\,K\bar{d}_n \rangle },
\label{e2_16e}
\ee
\be
\bar{d}_{n+1}= \bar{r}_{n+1}+\beta_n \bar{d}_n - 
 \frac{ \langle v_{n+1}, \, \bar{r}_{n+1}+\beta_n \bar{d}_n \rangle }
      { \langle v_{n+1},v_{n+1} \rangle } v_{n+1} \,.
\label{e2_16f}
\ee
\label{e2_16}
\esube
The definitions of $\bar{d}_0$ in \eqref{e2_16a} and $\bar{d}_{n+1}$ in \eqref{e2_16f}
ensure that
\be
\langle \bar{d}_n, v_n \rangle = 0
\label{e2_17}
\ee
at every iteration. Equation \eqref{e2_16c} ensures that $v_n=v+\tv_n$ where
\be
\langle \tv_n, v \rangle \To 0 \quad \mbox{as $n$ increases}\,.
\label{e2_18}
\ee
Equations (\ref{e2_16}b--d) entail the counterpart of \eqref{e2_02}:
\be
\langle \bar{r}_{n+1},\,\bar{d}_n \rangle = O(\tv_n^3)\,,
\label{e2_19}
\ee
and Eqs.~(\ref{e2_16}e,f) entail a counterpart of \eqref{e2_03}:
\be
\langle \bar{d}_{n+1},\,K\bar{d}_n \rangle = O(\tv_n^3)\,.
\label{e2_20}
\ee
The last three relations can be proved similarly to relations \eqref{e3_12} and \eqref{e3_13}.

\setcounter{equation}{0}
\renewcommand{\theequation}{A3.\arabic{equation}}
\section*{Appendix 3: \ Sample code of modified CGM for a two-component solitary
wave with prescribed propagation constants}

We will first present the code which uses the modified CGM \eqref{e2_14}, \eqref{e4_08}
to find a solitary wave of Eqs.~\eqref{e6_01}, then explain some of its steps, and, finally,
discuss how this code could be simplified. This code can be downloaded from \\
\verb+http://www.cems.uvm.edu/~lakobati/posted_papers_and_codes/code_in_Appendix3.m+.

\begin{verbatim}
% ----- Spatial and spectral domains:  ------------------------------------------
xlength=12*pi; ylength=12*pi;         % domain lengths along x and y
Nx=2^8;         Ny=Nx;                % number of points along x and y
dx=xlength/Nx;  dy=dx;                % mesh sizes along x and y
x=[-xlength/2:dx:xlength/2-dx]; y=x;  % domains along x and y
[X,Y]=meshgrid(x,y);                  % X and Y arrays of size Ny-by-Nx  
kx=2*pi/xlength*[0:Nx/2-1  -Nx/2:-1]; ky=kx;  % spectral domains along x and y
[KX,KY]=meshgrid(kx,ky);              % KX and KY arrays of size Ny-by-Nx 
DEL(:,:,1)=-(KX.^2+1*KY.^2); DEL(:,:,2)=DEL(:,:,1); % Fourier symbol of Laplacian 
% ----- Coefficients in the equation:  ------------------------------------------
mu(1)=7.89;  mu(2)=8.5;               % mu values        
W(:,:,1)=6*( (cos(X)).^2 + (cos(Y)).^2 ) - mu(1);
W(:,:,2)=6*( (cos(X)).^2 + (cos(Y)).^2 ) - mu(2); % potential minus mu  
F(1)=1; F12=0.5; F(2)=4;              % nonlinearity coefficients
Dt=0.9;                                 % Delta tau
% ----- Initial condition:             ------------------------------------------
u(:,:,1)= 0.8*exp(-1*(X.^2 + Y.^2)).*(1+0.1*X-0.2*Y);;
u(:,:,2)= 1.5*exp(-1*(X.^2 + Y.^2)).*(1+0.1*X-0.2*Y);;
% ----- Loop control variables:        ------------------------------------------
normDu=1;           % initialize the error norm to start the loop
normDu_accel=0.05;  % Acceleration begins when the error reaches this threshold;
                    % parameters c, b, gamma are computed until this threshold.
accelerate=0;       % this marker indicates that the acceleration has started
counter=0;          % counter of the number of iterations
% ----- Iterate until the error reaches the prescribed tolerance  ---------------
while normDu >= 10^(-10)
  counter=counter+1;
  if normDu >= normDu_accel & accelerate == 0    % compute N, E2, gammas  -------
    %  STEP 1:  Compute parameters of N:  b(2), c(1), c(2)  ~~~~~~~~~~~~~~~~~~~~~
    DELu=real( ifft2(DEL.*fft2(u)) );    usq=u.^2;
    for k=1:2
      L0u(:,:,k)=DELu(:,:,k)+(W(:,:,k)+F(k)*usq(:,:,k)+F12*usq(:,:,3-k)).*u(:,:,k); 
      u_L0u(k)=trapz(trapz( u(:,:,k).*L0u(:,:,k) ));  % <u, L0u> componentwise 
    end
    L11u1=L0u(:,:,1)+2*F(1)*usq(:,:,1).*u(:,:,1); L12u2=2*F12*u(:,:,1).*usq(:,:,2);
    L22u2=L0u(:,:,2)+2*F(2)*usq(:,:,2).*u(:,:,2); L21u1=2*F12*u(:,:,2).*usq(:,:,1);
    SumjLkjEj1(:,:,1)=L11u1+L12u2-L0u(:,:,1);
    SumjLkjEj1(:,:,2)=L21u1+L22u2-L0u(:,:,2);
    for k=1:2
      u_SumjLkjEj1(k)=trapz(trapz( u(:,:,k).*SumjLkjEj1(:,:,k) ));
      DELu_SumjLkjEj1(k)=trapz(trapz( DELu(:,:,k).*SumjLkjEj1(:,:,k) ));
      u_u(k)=trapz(trapz( u(:,:,k).*u(:,:,k) ));
      u_DELu(k)=trapz(trapz( u(:,:,k).*DELu(:,:,k) ));
      DELu_DELu(k)=trapz(trapz( DELu(:,:,k).*DELu(:,:,k) ));
    end
    b(1)=1;
    for k=1:2
      kappa(k)=( u_DELu(k)*DELu_SumjLkjEj1(k)-DELu_DELu(k)*u_SumjLkjEj1(k) )/...
               ( u_u(k)*DELu_SumjLkjEj1(k)-u_DELu(k)*u_SumjLkjEj1(k) );
    end
    b(2)=b(1)*(kappa(1)*u_u(1)-u_DELu(1)) * u_SumjLkjEj1(2) / ...
              ( (kappa(2)*u_u(2)-u_DELu(2)) * u_SumjLkjEj1(1) );
    c=b.*kappa;      
    u_Nu=c.*u_u-b.*u_DELu;   % <u, Nu> componentwise
    %  STEP 2: Compute eigenvector E2 = [rho2(1)*u1 rho2(2)*u2]^T  and  fft(N) ~~~
    rho2(1)=-u_Nu(2)/u_Nu(1);   rho2(2)=1;  
    rho1(1)=1;  rho1(2)=1;  % these coefficients of E1 are for uniform notations
    for k=1:2
      E1(:,:,k)=rho1(k)*u(:,:,k);    E2(:,:,k)=rho2(k)*u(:,:,k);
      fftN(:,:,k)=c(k)-b(k)*DEL(:,:,k);  % Fourier symbol of N componentwise
    end
    %  STEP 3: Compute gammas   ~~~~~~~~~~~~~~~~~~~~~~~~~~~~~~~~~~~~~~~~~~~~~~~~~~~
    E_NE(1)=sum( (rho1.^2).*u_Nu );  
    E_NE(2)=sum( (rho2.^2).*u_Nu ); 
    SumjLkjEj2(:,:,1)=rho2(1)*L11u1+rho2(2)*L12u2-L0u(:,:,1);
    SumjLkjEj2(:,:,2)=rho2(1)*L21u1+rho2(2)*L22u2-L0u(:,:,2);
    E_LE(1)=u_SumjLkjEj1(1)+u_SumjLkjEj1(2);
    E_LE(2)=sum( trapz(trapz( E2.*SumjLkjEj2 )) ); 
    lambda=E_LE./E_NE;   gamma=1+1./(lambda*Dt);
    %  STEP 4:  Update the solution:    ~~~~~~~~~~~~~~~~~~~~~~~~~~~~~~~~~~~~~~~~~~~
    E_L0u(1)=rho1(1)*u_L0u(1)+rho1(2)*u_L0u(2);
    E_L0u(2)=rho2(1)*u_L0u(1)+rho2(2)*u_L0u(2);
    u = u + Dt*( real( ifft2(fft2(L0u)./fftN) ) - ...
                 gamma(1)*E_L0u(1)/E_NE(1)*E1 - gamma(2)*E_L0u(2)/E_NE(2)*E2 );        
  else  % ----- use last computed N,E1,E2,lambda,gamma and start CGM  -------------
    if accelerate == 0
      accelerate = 1;
      for k=1:2
        L0u(:,:,k)=real( ifft2(DEL(:,:,k).*fft2(u(:,:,k))) ) + ...
                   (W(:,:,k)+F(k)*u(:,:,k).^2+F12*u(:,:,3-k).^2).*u(:,:,k); 
        u_L0u(k)=trapz(trapz( u(:,:,k).*L0u(:,:,k) ));  
        GAMMA(k)=1+1/lambda(k);
      end
      E_L0u(1)=rho1(1)*u_L0u(1)+rho1(2)*u_L0u(2);
      E_L0u(2)=rho2(1)*u_L0u(1)+rho2(2)*u_L0u(2);
      NE1=real( ifft2(fft2(E1).*fftN) );
      NE2=real( ifft2(fft2(E2).*fftN) );
      LL0u=L0u-GAMMA(1)*E_L0u(1)/E_NE(1)*NE1-GAMMA(2)*E_L0u(2)/E_NE(2)*NE2;
      r=real( ifft2(fft2(LL0u)./fftN) );    d=r;
    end
    for k=1:2
      Ld(:,:,k)=real( ifft2(DEL(:,:,k).*fft2(d(:,:,k))) ) + ...
                (W(:,:,k)+3*F(k)*u(:,:,k).^2+F12*u(:,:,3-k).^2).*d(:,:,k)+...
                2*F12*u(:,:,k).*u(:,:,3-k).*d(:,:,3-k);
    end
    E1_Ld=sum( trapz(trapz(E1.*Ld)) );   E2_Ld=sum( trapz(trapz(E2.*Ld)) );
    LLd=Ld-GAMMA(1)*E1_Ld/E_NE(1)*NE1-GAMMA(2)*E2_Ld/E_NE(2)*NE2;
    Nr_d=sum( trapz(trapz(LL0u.*d)) );   LLd_d=sum( trapz(trapz(d.*LLd)) );
    alpha = -Nr_d/LLd_d;
    u = u+alpha*d;
    for k=1:2
      L0u(:,:,k)=real( ifft2(DEL(:,:,k).*fft2(u(:,:,k))) ) + ...
                 (W(:,:,k)+F(k)*u(:,:,k).^2+F12*u(:,:,3-k).^2).*u(:,:,k); 
      u_L0u(k)=trapz(trapz( u(:,:,k).*L0u(:,:,k) ));     
    end
    E_L0u(1)=rho1(1)*u_L0u(1)+rho1(2)*u_L0u(2);
    E_L0u(2)=rho2(1)*u_L0u(1)+rho2(2)*u_L0u(2);
    LL0u=L0u-GAMMA(1)*E_L0u(1)/E_NE(1)*NE1-GAMMA(2)*E_L0u(2)/E_NE(2)*NE2;
    r = real( ifft2(fft2(LL0u)./fftN) );
    beta =  max(-sum( trapz(trapz(r.*LLd)) )/LLd_d, 0);
    d = r + beta*d;
  end
  normDu= norm(L0u(:,:,1))/norm(u(:,:,1))+norm(L0u(:,:,2))/norm(u(:,:,2));
  normDu_recorded(counter)=normDu;
end
figure(1); mesh(x,y,u(:,:,1)); figure(2); mesh(x,y,u(:,:,2)); 
figure(3); plot([1:counter],log10(normDu_recorded))
\end{verbatim}

Step 1 inside the while-loop implements Eqs. (4.12) and (4.13) (see also (4.21)) of \cite{gP}
using the notations of that paper. For example, \verb+DELu+,
\verb+L0u+, \verb+SumjLkjEj1+ in the code stand, respectively, for 
$\nabla^2 {\bf u}_n$, ${\bf L}_0{\bf u}_n$, $\sum_{j=1}^2 L_{kj}e_{j1}$; note
that all these quantities are two-dimensional vectors. Notations involving the
underscore denote inner products; for example, \verb+u_L0u(k)+ denotes
$\langle u^{(k)}_n, ({\bf L}_0{\bf u})^{(k)}_n \rangle$, $k=1,2$, where the superscript
$(k)$ means the $k$th component of the two-dimensional vector.

Step 2 implements Eqs. (4.15)--(4.18) of \cite{gP}. Here \verb+rho1(k)+ denotes
$\rho^{(1,k)}$ in the notations of Eq.~\eqref{e4_06} of {\em this} paper; note that,
for the convenience of coding, 
the order of the superscripts is reversed compared to \cite{gP}.

Step 3 implements Eqs. (4.5) of \cite{gP}. Note that $\alpha_k$ of \cite{gP} is
denoted as $\lambda^{(k)}$ in this paper. Also note that in the code, quantities
like \verb+E_NE(1)+ denote $\langle {\bf e}^{(1)}, \, {\bf N\,e}^{(1)} \rangle$, i.e. here,
the superscript refers to the particular eigenvector ${\bf e}^{(k)}$ of ${\bf N}^{-1}{\bf L}$.

Step 4 implements Eq. (4.19) of \cite{gP}.

At the first iteration of the CGM, in addition to implementing Eqs.~\eqref{e2_14a}
of this paper, one also computes $\Gamma^{(k)}$ and ${\bf N e}^{(k)}$, $k=1,2$,
so that these quantities are {\em not} computed again at subsequent iterations of the 
CGM. Notations \verb+LL0u+ and \verb+LLd+ stand for ${\bf L^{(0,\rm mod)} u}_n$ and
${\bf L^{(\rm mod)} d}_n$. The remainder of the code implements Eqs.~\eqref{e2_14} and
\eqref{e4_08} of this paper. 

\smallskip

Note that the CGM iterations start when ${\bf N}$, ${\bf e}^{(k)}$, and $\Gamma^{(k)}$ are
found rather imprecisely: accuracy (i.e., the value of \verb+normDu_accel+) of 5\% 
is used in the above code and in the examples
reported in Section 6, and we verified that the code still worked when this accuracy was 
lowered to as much as 10\%. Moreover, even if ${\bf N}$ etc. were computed up to a higher
accuracy, the functions ${\bf e}^{(k)}$ could still satisfy the eigenrelations \eqref{e4_04} only
approximately. While this approximation is very close (99\% in the least-squares sense)
for ${\bf e}^{(1)}$, it is only about 70\% for ${\bf e}^{(2)}$; see the second column in
Table 1 of \cite{gP}. These considerations suggest that the code could still work if the
${\bf N}$ given by the expression above Eq.~\eqref{e4_03} is replaced by a simpler 
expression \eqref{e6_11}.
The constant $c$ there should be computed by a straightforward generalization
of Eq. (3.11) of \cite{gP}, whereby the lines in Step 1 of the above code 
starting with the second for-loop through the end of that step are replaced by:

\begin{verbatim}
for k=1:2
  u_u(k)=trapz(trapz( u(:,:,k).*u(:,:,k) ));       % <u, u>  componentwise
  u_DELu(k)=trapz(trapz( u(:,:,k).*DELu(:,:,k) )); % <u, DELu>  componentwise
end
DELu_DELu=sum( trapz(trapz( DELu.*DELu )) );  % <DELu, DELu> as a scalar, etc
u_SumjLkjEj1=sum( trapz(trapz( u.*SumjLkjEj1 )) );
DELu_SumjLkjEj1=sum( trapz(trapz( DELu.*SumjLkjEj1 )) );
c = ( u_SumjLkjEj1*DELu_DELu - DELu_SumjLkjEj1*sum(u_DELu) )/...
    ( u_SumjLkjEj1*sum(u_DELu) - DELu_SumjLkjEj1*sum(u_u) );    
u_Nu=c*u_u-u_DELu;        % <u, Nu> componentwise
\end{verbatim}

Note that the reason why the inner products \verb+u_u(k)+ and \verb+u_DELu(k)+
are computed for each component of ${\bf u}_n$ rather than for the entire vector
function, as \verb+DELu_DELu+ etc, is that they are needed to compute the individual
components of \verb+u_Nu+. The latter are needed to compute $\rho^{(2,1)}$ in Step 2.

We did not test the performance of such a  simplified code because the goal of this paper
is the {\em proposal} of modified CGMs for solitary waves and {\em not optimization}
of those methods.

\newpage

\hfill {\bf T.I. Lakoba}

\vfill

\newpage

\hfill {\bf T.I. Lakoba}\\

\vfill

\newpage

\hfill {\bf T.I. Lakoba}\\

\vfill

\end{document}